\documentclass[amsmath,nobibnotes,aps,amssymb, prx,showpacs,superscriptaddress,twocolumn, longbibliography, reprint,nofootinbib]{revtex4-2}
\usepackage[colorlinks=true,linkcolor=blue,urlcolor=blue,citecolor=blue,pdfusetitle]{hyperref}
\usepackage{hyperref}
\usepackage{graphicx}
\usepackage{dcolumn}
\usepackage{bm}
\usepackage{physics}
\usepackage{bbm}
\usepackage[T1]{fontenc}
\usepackage[utf8]{inputenc}
\usepackage{setspace}
\usepackage{textcase}
\usepackage{chngcntr}
\usepackage{xcolor}
\usepackage{xcolor}
\definecolor{color1}{HTML}{1E88E5}

\makeatletter
\let\orig@seccntformat\@seccntformat
\renewcommand{\appendix}{%
    \par
    \setcounter{section}{0}%
    \setcounter{subsection}{0}%
    \gdef\thesection{SM. \Roman{section}}%
    \gdef\thesubsection{SM. \Roman{section}.\Alph{subsection}}%
    \providecommand{\theHsection}{}%
    \providecommand{\theHsubsection}{}%
    \gdef\theHsection{SM.\Roman{section}}%
    \gdef\theHsubsection{SM.\Roman{section}.\Alph{subsection}}%
    \gdef\@seccntformat##1{%
        \ifx##1\section
        SM \Roman{section}.\quad
        \else
        \ifx##1\subsection
        \Alph{subsection}.\quad
        \else
        \orig@seccntformat{##1}%
        \fi
        \fi
    }%
    \renewcommand{\section}{%
        \@startsection{section}{1}{\z@}%
        {-6.0ex \@plus -0.6ex \@minus -0.2ex}%
        {6.0ex \@plus 0.3ex}%
        {\normalfont\normalsize\bfseries\centering}%
    }%
    \renewcommand{\subsection}{%
        \@startsection{subsection}{2}{\z@}%
        {-3ex \@plus -0.4ex \@minus -0.2ex}%
        {3ex \@plus 0.2ex}%
        {\normalfont\normalsize\bfseries\centering\NoCaseChange}%
    }%
    \setcounter{figure}{0}%
    \renewcommand{\thefigure}{S\arabic{figure}}%
    \providecommand{\theHfigure}{}%
    \gdef\theHfigure{S\arabic{figure}}%
    \setcounter{table}{0}%
    \renewcommand{\thetable}{S\arabic{table}}%
    \providecommand{\theHtable}{}%
    \gdef\theHtable{S\arabic{table}}%
    \setcounter{equation}{0}%
    \renewcommand{\theequation}{S\arabic{equation}}%
    \providecommand{\theHequation}{}%
    \gdef\theHequation{S\arabic{equation}}%
    \setcounter{page}{1}%
    \renewcommand{\thepage}{S\arabic{page}}%
}

\makeatother

\begin{document}
    \title{High-Probability Heralded Entanglement via Repeated Spin-Photon Phase Encoding with Moderate Cooperativity}    
    \author{Yu Liu}
    \email{yu.liu@uni-ulm.de}
    \affiliation{Institut f\"{u}r Theoretische Physik, Universit\"{a}t Ulm, Albert-Einstein-Allee 11, 89069 Ulm, Germany}
    \affiliation{Center for Integrated Quantum Science and Technology (IQST), 89081 Ulm, Germany}
    \author{Martin B Plenio}
    \email{martin.plenio@uni-ulm.de}
    \affiliation{Institut f\"{u}r Theoretische Physik, Universit\"{a}t Ulm, Albert-Einstein-Allee 11, 89069 Ulm, Germany}
    \affiliation{Center for Integrated Quantum Science and Technology (IQST), 89081 Ulm, Germany}
    
    \date{\today}
    
    \begin{abstract}
        We propose a heralded high-probability scheme to generate remote entanglement between moderate-cooperativity spin-cavity registers with high fidelity. In conventional single-shot interfaces, limited cooperativity restricts the spin-conditional optical response and thus strongly suppresses the success probability. Our proposal instead recycles a single incident photon for repeated interactions with the spin-cavity register, such that a small spin-conditional phase shift acquired on each round trip accumulates coherently to enable remote entanglement. Moreover, the repeated scheme enables higher spin-photon encoding efficiency by using a spectral-width-scaling photon pulse with a shorter duration. We show that, for realistic imperfections and losses, this repeated phase-encoding approach produces high-fidelity entangled states with an appreciable success probability even at cooperativity $C\sim1$. Our protocol is particularly well suited to weakly coupled, cavity-based solid-state spin platforms and provides a route toward hybrid, photon-loss-tolerant distributed quantum computing.
    \end{abstract}
    \maketitle
    {\it Introduction.---}
    Scalable quantum networks~\cite{Kimble2008, Wehner2018} demand high-rate, high-fidelity entanglement between remote modular matter-qubit registers~\cite{Bose1999, Barrett2005, Lim2005, Loock2006, Stephenson2020, Li2024}, typically mediated by a flying photonic qubit, toward distributed quantum computing~\cite{CiracEH99,EisertJPP00, Jiang2007}. Optical heralded entanglement protocols based on the single-photon interference and spin-dependent scattering have been demonstrated to entangle distant spin-cavity interfaces, like color centers~\cite{Nguyen2019, Stas2022, Knaut2024, Bhaskar2020}, trapped ions~\cite{Krutyanskiy2023, Ruskuc2025}, quantum dots~\cite{ Delteil2016, Heindel2023}, superconducting~\cite{CampagneIbarcq2018}, atoms~\cite{Ritter2012, Daiss2021, vanLeent2022} and form the basis for distributed quantum information processing. 
    
    In spin-cavity interfaces, the cavity enhances the effective spin-photon interaction through multiple {\it internal cycles} of a photon, which can lead to a large cooperativity $C$~\cite{Reiserer2015}. Specifically, in reflection-based schemes, $C\gg1$ enables a single spin to exert well-defined conditional control over an incident photon by coherently shaping the complex reflection amplitude and phase. In phase encoding, the spin imprints a state-dependent phase shift on the reflected photon, ideally approaching $\pi$~\cite{Duan2004, Hu2008, Bonato2010}. 
    In amplitude encoding, the spin yields a unit reflection in one spin state and a vanishing one in the other~\cite{Nemoto2014, Koshino2012, Bhaskar2020}. Increasing $C$ reduces incoherent loss, routing more photons into the desired reflected mode to boost the success probability, while also potentially improving the heralded fidelity~\cite{Beukers2024}.
    Moreover, in the Purcell regime, large cooperativity leads to emission predominantly into the cavity mode and a single propagating output, enabling efficient single-photon sources and spin-photon interfaces~\cite{Riedel2017, Wang2025Purcell, Yurgens2024}.

    Despite these attractive advantages, achieving and stabilizing high cooperativity is technically demanding, especially in solid-state systems, due to homogeneous broadening~\cite{Lodahl2015, Weiss2025}, small spin-photon coupling~\cite{Riedel2017, Katsumi2025}, and limited cavity finesse. As a result, the coherent reflection competes with spontaneous decay, and the heralding probability is inherently significantly suppressed in the moderate-cooperativity regime (MCR) $C\sim1$~\cite{Omlor2025}.
    
    To address the low success probability in the MCR, we introduce an entangling protocol that employs additional {\it external cycles} to recycle a single photon for repeated spin-photon phase encoding. This protocol works in a far-detuned dispersive regime, where the incident photon is predominantly reflected while acquiring a small spin-dependent phase shift~\cite{Koshino2012}. 
    Routing the photon back to the cavity through the external cycles allows accumulating phase coherently, reaching a conditional phase difference of $\pi$ after multiple rounds. 
    Off-resonant operation strongly suppresses spin excitation and the associated incoherent decay, improving photon survival and yielding a probability that approaches 1. Additionally, with suitable optimization, it retains a high success probability under realistic loss and imperfections, and enhances the entanglement generation rate with a shorter incident pulse employed without relying on high cooperativity.
    
    \begin{figure}[t]
        \centering
        \includegraphics[width=0.4\textwidth]{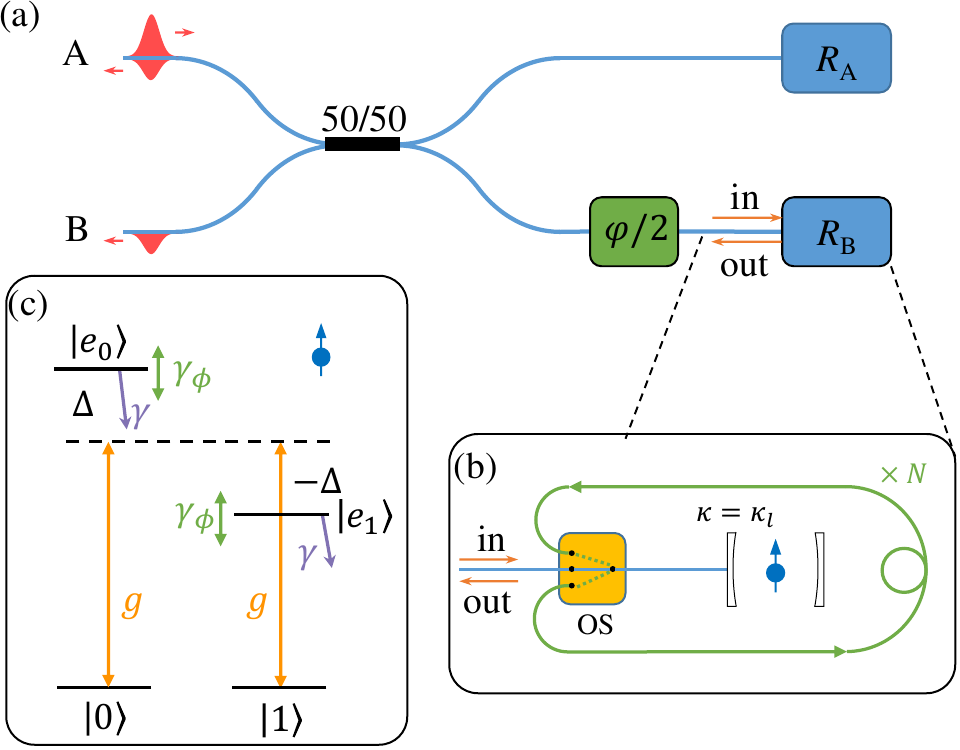}
        \caption{The heralded entangling based on repeated spin-photon phase encoding. (a) The schematic view of the entangling circuit, in which two registers $R_A$ and $R_B$ are connected by a Mach-Zehnder interferometer. Here, the upper side is labeled $A$. A single photon is sent from port $A$ and splits at a $50/50$ beam splitter (BS), and the conditional phase encoding is implemented with the corresponding register. After the encoding, the reflected photons are routed back and combined at BS. The remote entanglement is generated when one photon reaches the port $A$ or $B$ and the detector clicks. (b) A local register exhibiting a four-level system and an optical cavity and an additional optical cycle (green curve). (c) A single cavity mode couples with two optical transitions at a strength $g$.}
        \label{fig_scheme}
        \vspace{-0.2cm}
    \end{figure}
    
    {\it Repeated spin-photon phase encoding.---}
    We consider a local register hosting a four-level spin system placed in a single-sided optical cavity, as shown in Fig.~\ref{fig_scheme}(a), where the cavity mode couples with two identical optical transitions. We denote by $\mathcal{T}_s$ with $s\in\{0,1\}$ the optical transitions between the ground state $|s\rangle$ and the excited state $|e_s\rangle$. 
    The matter qubit is encoded in the ground-state subspace and initialized to the superposition state $(|0\rangle + |1\rangle)/\sqrt{2}$ in both registers.
    In our proposal, heralded entanglement between two registers is generated via a state-dependent reflected photon, with the central frequency of the incident photon resonant with the cavity $(\omega_p=\omega_c)$.
    The entangling setup is shown in Fig.~\ref{fig_scheme}(a), in which two registers $R_A$ and $R_B$ are connected by a Mach-Zehnder interferometer (MZI). 
    
    In accordance with the input-output formalism for a cavity coupled to a single optical transition, an incident single-photon $\int d\omega\tilde{u}(\omega)a^\dagger(\omega)|0\rangle$, where $\omega$ denotes the detuning between the incident photon and the cavity, undergoes a spin-conditioned reflection. For the spin state $|s\rangle$, the reflection maps the envelope as $\tilde{u}(\omega)\rightarrow r_s(\omega)\tilde{u}(\omega)$, with reflection coefficient $r_s(\omega)=-R_s(\omega)\exp[i\theta_s(\omega)]$, and $r_s(\omega)$ is given by~\cite{Omlor2025, Reiserer2015}
    \begin{align}
        r_s(\omega) = 1- \frac{2\kappa_l \left[\gamma + 2i(\Delta_s - \omega)\right]}{(\kappa-2i\omega)[\gamma+2i(\Delta_s-\omega)]+4g^2},
        \label{eq_r_i}
    \end{align}
    where $\kappa=\kappa_l$ is the loss rate of a single-sided cavity; $\gamma = \gamma_{\mathrm{se}}+2\gamma_\phi$ represents the total decay rate, incorporating spontaneous emission and pure dephasing; $\Delta_s$ corresponds to the detuning between  $\mathcal{T}_s$ and the cavity, and their values are tuned symmetrically as $\Delta_0=-\Delta_1=\Delta$ to make $r_0(\omega) \approx r_1^*(\omega)$ near $\omega=0$. The cooperativity is defined as $C=4g^2/(\kappa\gamma)$.
    
    The entangling process begins with a single photon $\hat{a}^\dagger_A$ injected into a $50/50$ beam splitter (BS). At the BS, the photon creation operator transforms according to $\hat{a}_A^\dagger\rightarrow(\hat{a}_B^\dagger+i\hat{a}_A^\dagger)/\sqrt{2}$ and $\hat{a}_B^\dagger\rightarrow(i\hat{a}_B^\dagger+\hat{a}_A^\dagger)/\sqrt{2}$, thereby generating a coherent superposition of paths $A$ and $B$. The photon is then sent to interact with the corresponding register and be spin-conditionally reflected and modulated. The reflected photons are then routed by a 3$\times$1 optical switch(OS) back into the cavity along the external cycle (green path in Fig.~\ref{fig_scheme}(b)), so that the photons interact with the register again. After repeating this cycle $N$ times, the register modifies the input spectral envelope as follows:
    \begin{align}
        \left[|0\rangle\langle0| r_0(\omega)^N + |1\rangle\langle1| r_1(\omega)^N\right] \tilde{u}(\omega).
    \end{align}
    After the $N$-th round, the OS routes the photon back to the BS. Provided that the entangling condition $r_+(\omega)=0$ holds, corresponding to a $\pi$ conditional phase difference, a single-photon detection event at either output port (A or B) heralds a remote spin-spin Bell state. Here $2r_{\pm}(\omega)=r_0(\omega)^N \pm r_1(\omega)^N$. 
    To maximize the reflection probability $\int d\omega|r_s(\omega)^N\tilde{u}(\omega)|^2$, we operate at large detuning $\Delta\gg\gamma$. Off-resonant operation strongly suppresses the excited-state population and the associated decoherence $\sim\gamma p_e$, which is key to achieving a high success probability at moderate cooperativity.
    
    {\it Heralded entangling generation.---} For a monochromatic pulse or long pulse (i.e.,~$\tilde{u}(\omega)=\delta(\omega)$), the reflection coefficient Eq.~\eqref{eq_r_i} is given by $r_0=-Re^{i\theta}$. 
    For large detuning, $\theta\approx C\gamma/\Delta$ and $R\approx 1-\theta^2/(2C)$. After $N$ rounds, the state leading to clicks at the detectors is
    \begin{align}
        \frac{\hat{a}_A^\dagger}{\sqrt{2}}  \left( r_+ |\Phi^+\rangle + r_- |\Phi^-\rangle +r_+|\Psi^+\rangle\right)+\frac{i\hat{a}_B^\dagger}{\sqrt{2}} r_-|\Psi^- \rangle\, .
    \end{align}
    A click at port $B$ heralds a perfect Bell state $|\Psi^-\rangle=(|01\rangle - |10\rangle)/\sqrt{2}$ with probability $P_B^{(\delta)} = R^{2N}\sin^2(N\theta)/2$ and unit fidelity $F_B^{(\delta)} = 1$. Here $|mn\rangle = |m\rangle_A|n\rangle_B$. By contrast, under the entangling condition, one click event at port $A$ heralds an entangled state $|\Phi^-\rangle = (|00\rangle -|11\rangle)/\sqrt{2}$ with probability $P_A^{(\delta)} = R^{2N}[1+\cos^2(N\theta)]/2$ and fidelity $F_A^{(\delta)}=\sin^2(N\theta)/[1+\cos^2(N\theta)]$. 
    To fulfill the entangling condition, we choose the detuning such that the spin-dependent relative phase $\theta$ accumulates linearly per cycle and reaches $\pi$ after $N=\pi\Delta/(2C\gamma)$ rounds.
    Note that the total success probability $P_t^{(\delta)} = R^{2N}$, which approaches $1$ with $\Delta\rightarrow\infty$~(See~\ref{Sec_SI_reflcetion_based}). Note that, in practice, with finite losses and spin coherence, one optimal success probability is obtained at a finite $N$, as shown in Fig.~\ref{fig_optimization}(b).
    \begin{figure}[t]
        \centering
        \includegraphics[width=0.48\textwidth]{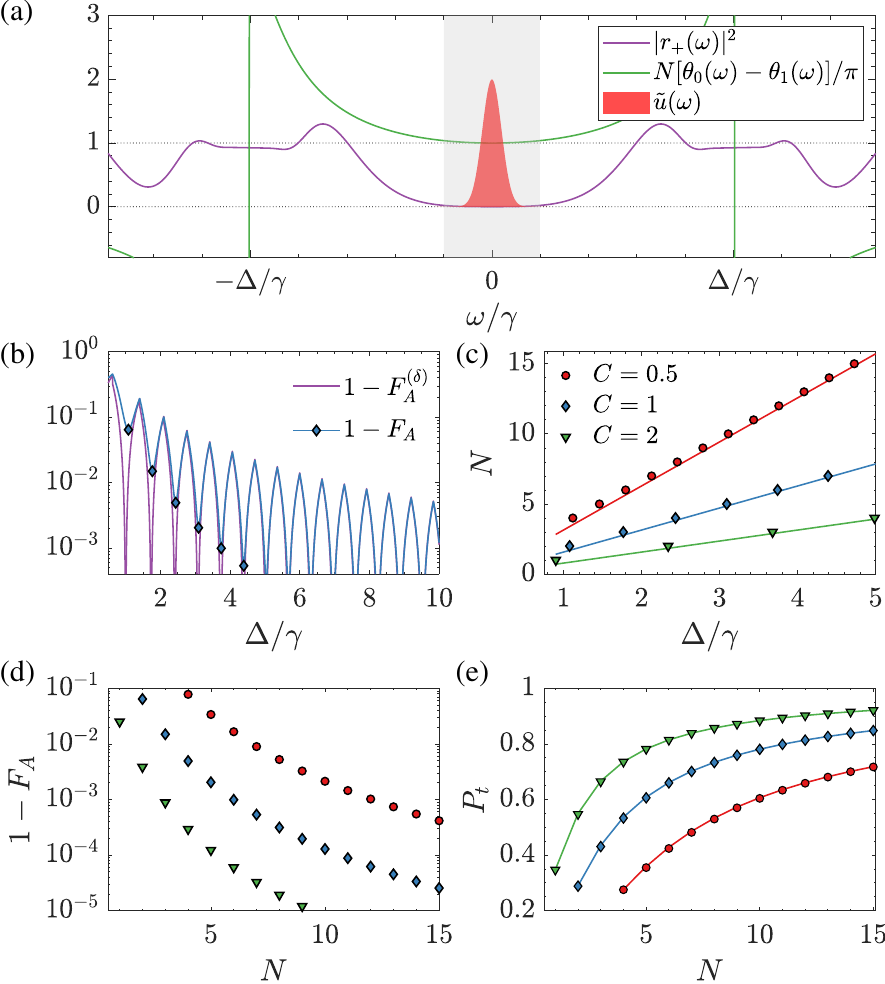}
        \caption{Simulation of the entangling proposal. (a) The input Gaussian spectral envelope $\tilde{u}(\omega)$ is centered in the entangling window (gray region) around $\omega/\gamma=0$, where $|r_+(\omega)|^2\approx0$ (purple line) and $N(\theta_0-\theta_1)\approx\pi$ (green line). Parameters: $\Delta/\gamma=6.2826$, $N=5$, spectral width $\sigma_w =\gamma/5$ and cooperativity $C=1$. (b) Simulated infidelities for the monochromatic pulse, $1-F_A^{(\delta)}$ (purple line), and for a Gaussian pulse, $1-\tilde{F}_A$ (blue line), as functions of the detuning $\Delta/\gamma$ with $C=1$. The blue squares denote the optimal detuning values $\Delta$ to maximize the fidelity. (c) The required repetition number $N$ as a function of $\Delta$ for $C=0.5,1,2$, shown as red circles, blue squares, and green triangles, respectively, which lie very close to the monochromatic result $N=\pi\Delta/(2C\gamma)$ (solid lines). (d,e) The simulated infidelities $1-\tilde{F}_A$, total success probabilities $P_t$ as a function of the repetition $N$ for varying $C$ of a Gaussian pulse. The solid lines in (e) show the corresponding monochromatic probabilities. Parameters: $\kappa/\gamma=200$ throughout this paper.}
        \vspace{-0.2cm}
        \label{fig_simulation}
    \end{figure}
    
    In practice, we consider a single incident photon with spectral envelope, described by a Gaussian envelope $\tilde{u}(\omega)=(\pi\sigma_\omega^2)^{-1/4}\exp(-\omega^2/2\sigma_\omega^2)$, with the normalization $\int d\omega|\tilde{u}(\omega)|^2=1$. The detection probabilities at the two ports are $P_A=\int d\omega[|2r_+(\omega)|^2+|r_-(\omega)|^2]|\tilde{u}(\omega)|^2/2$ and $P_B=\int d\omega|r_-(\omega)|^2|\tilde{u}(\omega)|^2/2$, with fidelities $F_B=1$ and $F_A=P_B/P_A$. 
    As illustrated in Fig.~\ref{fig_simulation}(a), we choose a highly localized photon with a fixed bandwidth $\sigma_\omega=\gamma/5\ll\Delta$, so that its spectral envelope lies entirely within the entangling window (gray region), where $N[\theta_0(\omega)-\theta_1(\omega)]\approx\pi$ (green curve) and $|r_+(\omega)|^2\approx0$ (purple curve).  For a Purcell regime $C=1$, we set $\kappa/\gamma=200$ here and after; the resulting fidelity $F_A$ is shown as the blue solid line in Fig.~\ref{fig_simulation}(b), and the blue squares indicate the optimal detuning $\Delta$ that maximizes $F_A$. The purple line shows the simulated fidelity $F_A^{(\delta)}$, which serves as a lower bound for the Gaussian pulse. The required repetitions $N$ for $C=0.5,1,2$ (red circles, blue squares, and green triangles) are shown in Fig.~\ref{fig_simulation}(c) and are close to the monochromatic limit (solid lines in the same colors). 
    In Fig.~\ref{fig_simulation}(d,e), we show the corresponding simulated infidelities $1-F_A$ and probabilities $P_t$ as a function of $N$. Note that increasing $C$ requires a larger $\Delta$ to achieve the same phase increment $\theta$, which broadens the entangling window bounded by $\mathrm{Im}[r_s(\omega)]=0$ at $\omega=\pm\Delta$ and further improves the fidelity. The optimal probabilities $P_t$ closely follow the value $P_t^{(\delta)}$ (solid lines), indicating that high fidelity can be achieved without reducing the success probability.
    
    \begin{figure}[t]
        \centering
        \includegraphics[width=0.48\textwidth]{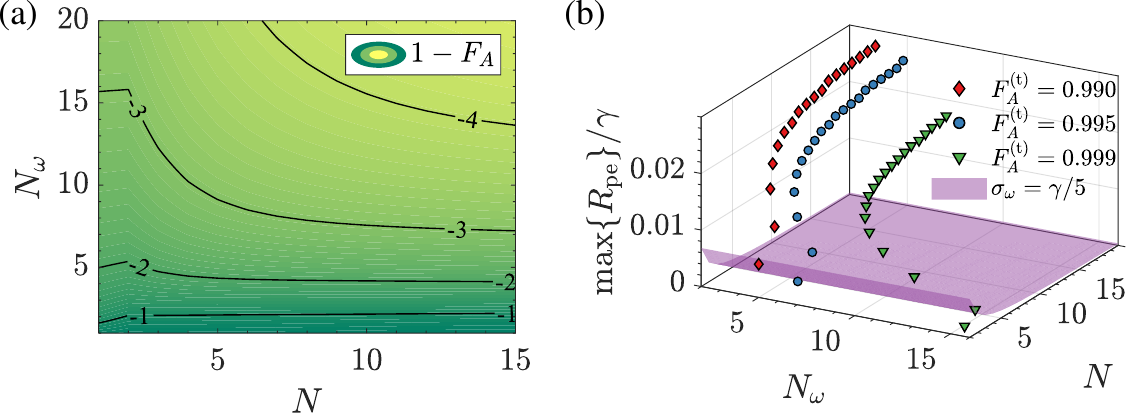}
        \caption{(a) The simulated infidelity with different spectral width $\sigma_\omega = \Delta/N_\omega$ at each $N$. (b) Maximal encoding rates for fidelity thresholds $F_A^{(\mathrm{t})}=0.99$, $0.995$, and $0.999$ are shown by red diamonds, blue circles, and green triangles, respectively, where each point uses the optimal $N_\omega$ for a given $N$. The width-fixed result $\sigma_\omega=\gamma/5$ is indicated by the purple $N_\omega$-independent surface. Parameters: $N_t=10$, $C=2$.}
        \label{fig_width_varying}
        \vspace{-0.2cm}
    \end{figure}
    {\it Increasing the phase encoding rate.---}
    A shorter pulse to increase the phase-encoding rate $R_{\mathrm{pe}}=P_t/T_{\mathrm{pe}}$ is possible by using a width-scaling pulse, where $P_t$ is the heralding probability per single-photon attempt and $T_{\mathrm{pe}}=N N_t/\sigma_\omega$ is the pure phase-encoding duration. Here $N_t$ is the number of temporal widths $\sigma_t=1/\sigma_{\omega}$. For example, choosing $\sigma_\omega=\Delta/N_\omega$ makes $T_{\mathrm{pe}}=N_tN_\omega\pi/(2C\gamma)$ independent of $N$ for a given $N_\omega$. Note that $R_{\mathrm{pe}}$ captures the phase-encoding-limited efficiency per attempt and hence upper-bounds the wall-clock entangling rate, as non-encoding steps increase the time and additional loss reduces the success probability. As shown in Fig.~\ref{fig_width_varying}(a), increasing $N_\omega$ improves the fidelity, but it also increases $T_{\mathrm{pe}}$ and thus lowers the encoding rate, implying a fidelity-rate trade-off. We therefore define a fidelity-constrained encoding rate for port-$A$ heralding, requiring the fidelity to exceed a threshold $F_A^{(\mathrm{t})}$. 
    In Fig.~\ref{fig_width_varying}(b), we plot the maximum rates for the thresholds $F_A^{(\mathrm{t})}=0.99,\,0.995,$ and $0.999$ (red diamonds, blue circles, and green triangles), obtained by optimizing over $N_\omega$ for each fixed $N$. Compared with the width-fixed scheme $\sigma_\omega=\gamma/5$ (purple surface), the width-scaling strategy yields higher rates. For instance, for $F_A^{(\mathrm{t})}=0.99$ at $N=5$, we obtain $R_{\mathrm{pe}}/\gamma=0.023$ at $N_\omega\approx4.36$, corresponding to a $7.36$ times speedup.
    
    {\it Protocol optimization.---} 
    We next discuss the extensions and optimization of the protocol to realistic imperfections and applications. Unless otherwise stated, our simulations use the width-fixed scheme with $\sigma_\omega=\gamma/5$.
    The first imperfection of the system is parameter disorder among the four transitions $\mathcal{T}_{s,M}$, which may arise from the fabrication. It leads to four distinct reflection coefficients $r_{s,M}(\omega)$ with $M\in\{A,B\}$. For each register $R_M$, the entangling condition becomes $r_{+,M}(\omega)=0$, which can be satisfied by appropriately tuning the detuning values $\Delta_{s,M}$. More importantly, the residual mismatch between the reflections associated with $R_A$ and $R_B$ can be compensated by applying an additional correction operation $U(\sqrt{r},\varphi/2)=\sqrt{r}e^{i\varphi/2}$ (green blocks in Fig.~\ref{fig_scheme}(a)) on the larger-reflection path (here, e.g., path $B$), where $r=|r_{0,A}(0)/r_{0,B}(0)|^N$ and $\varphi=N[\theta_{0,A}(0)-\theta_{0,B}(0)]$ (see~\ref{Sec_SI_register_difference}). This correction is calibrated at $\omega=0$ and thus only approximately compensates over the finite bandwidth, working best for the narrowband pulses used here. And the repetition numbers of both registers are identical to make sure that the reflected photons get to the BS at the same time. To show this, we set different coupling strengths such that the four cooperativities are $1.5,2,2.5,3$. The simulated uncorrected infidelities $1-F_{A}$ and $1-F_{B}$ are shown as red squares and orange circles in Fig.~\ref{fig_optimization}(a), respectively. While the corrected infidelities $1-F_{A}^{(\mathrm{c})}$ and $1-F_{B}^{(\mathrm{c})}$ are shown as blue and green triangles, demonstrating the effectiveness and improvement of the correction. For comparison, we also plot the ideal reference infidelity $1-F_A^{(\mathrm{id})}$ (purple line) for four identical transitions with cooperativity $C=2$.
    
    \begin{figure}[t]
        \centering
        \includegraphics[width=0.48\textwidth]{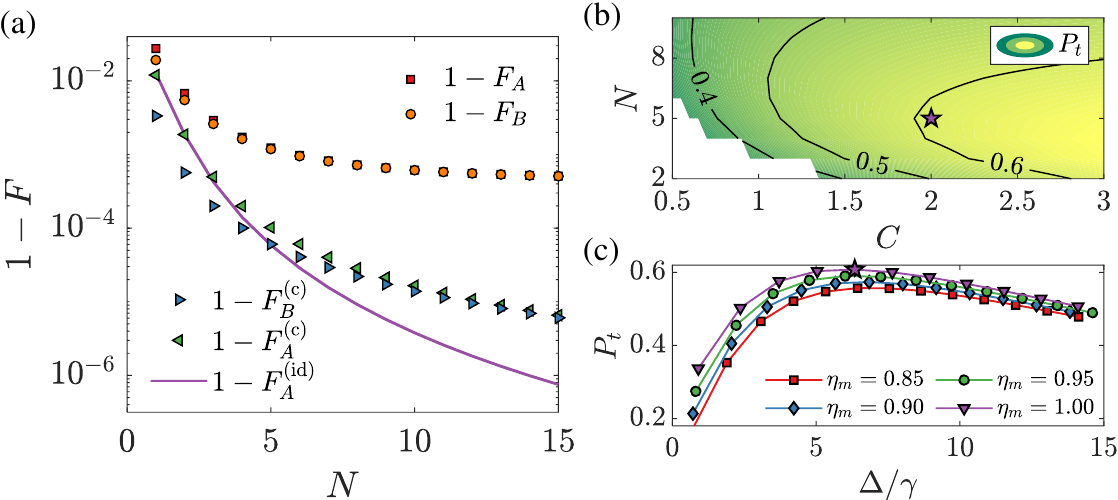}
        \caption{(a) The simulated infidelities for four transitions with cooperativity values $C=1.5,\,2,\,2.5,\,3$. The infidelities $1-F_{A(B)}^{(c)}$, denoted by left(right)-pointing green(blue) triangles, are improved compared to the uncorrected infidelity $1-F_{A(B)}$ denoted by red squares(orange circles), approaching the infidelity with four identical transitions with $C=2$ indicated by the purple line. (b) The success probability $P_t$ versus cooperativity $C$ and repetition $N$ with the blank region having $F_A<0.99$. The efficiencies are $\eta_r,\eta_i=0.9886,\,0.99$. (c) The probability $P_t$ achieved with the correction $U_{\pi}$ for different overlaps $\eta_m$ with the same efficiencies in (b) at $C=2$.}
        \label{fig_optimization}
        \vspace{-0.2cm}
    \end{figure}
    
    One other imperfection is the loss of photons, including intrinsic cavity loss $\kappa_i$ and losses in the optical circuit, characterized by the efficiencies $\eta_i=1-\kappa_i/\kappa$ and $\eta_r$. For the monochromatic pulse, the reflection $r_s(\omega)$ is effectively scaled as $\eta_i^2 \eta_r^{1/2} r_s(\omega)$ (see~\ref{Sec_SI_loss}), so that the success probability is $\eta_i^{4N}\eta_r^{N} P_t$. 
    Consequently, at a fixed loss level, the success probability is maximized at an optimal repetition number $N$ that balances reduced spontaneous emission at larger detuning against accumulated loss of photons. Accordingly, the $P_t$, with fidelity $F_A>0.99$, peaks at a finite, cooperativity-dependent $N$, as presented in Fig.~\ref{fig_optimization}(b). For the efficiencies $\eta_r=0.9886$ and $\eta_i=0.99$, we obtain $P_t\sim0.5$ for $C>1$.
    
    The next imperfection we account for is spatial mode mismatch in the fiber-based Fabry-Pérot cavity (FFPC), which is widely used in the implementation of a solid-state spin-based spin-photon interface. The incident photon carries a single-mode fiber(SMF) mode that can be decomposed as $|f\rangle=\sqrt{\eta_m}|m\rangle+\sqrt{1-\eta_m}|n\rangle$. Here, the matched mode $|m\rangle$ enters the cavity and acquires the spin-dependent reflection $r_s(\omega)$, whereas the unmatched mode $|n\rangle$ is directly reflected with a reflection coefficient $r_ne^{i\theta_n}$. Here, $\eta_m$ quantifies the overlap between the modes of SMF and FFPC, $\theta_n=0$ and $r_n=1$. The relevant effective reflection is $\tilde{r}_s(\omega)=\eta_m r_s(\omega)+(1-\eta_m) r_n e^{i\theta_n}$~(see~\ref{Sec_mode_mismatch}). Consequently, the amplitude $|\tilde{r}_s(\omega)|$ is strongly reduced by the destructive matched-unmatched interference upon projection of the photon back into the SMF~\cite{Gallego2016, Bick2016}. This SMF-filtering loss can be mitigated in two ways. The first is increasing $\eta_m$ by using mode-matching components and $\eta_m>0.9$ has been achieved~\cite{Wang2025,Gao2019,Gulati2017}. Secondly, a selective correction $U_{\pi}=|m\rangle\langle m|-|n\rangle\langle n|$ (making $\theta_n=\pi$) is employed, which converts the unmatched modes into a useful entangling resource like a matched one. This correction maintains high fidelity across the different optimal detuning values shown in Fig.~\ref{fig_optimization}(c), and the difference in $P_t$ for varying $\eta_m$ is small and vanishes at large $\Delta/\gamma$.
    
    \begin{figure}[t]
        \centering
        \includegraphics[width=0.48\textwidth]{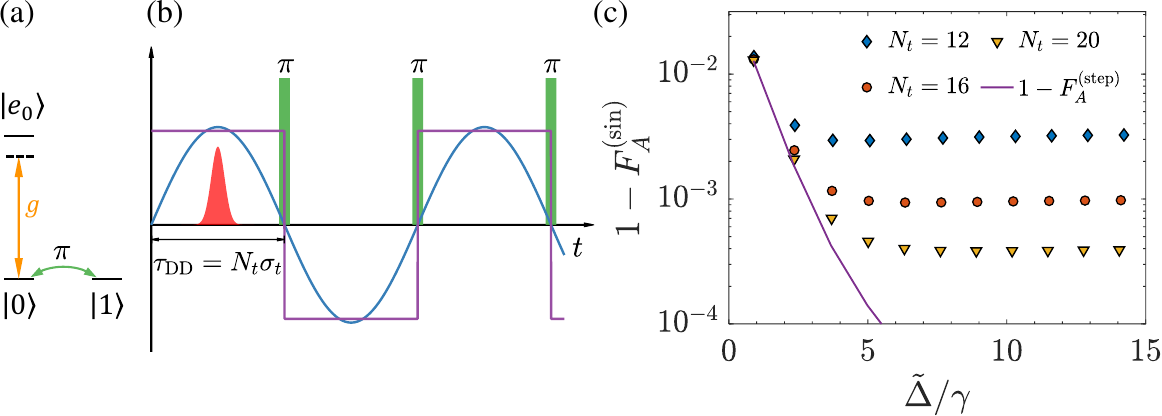}
        \caption{(a) Three-level system. (b) Dynamical-decoupling-based phase encoding. The $\pi$ pulses (green blocks) periodically flip $|0\rangle$ and $|1\rangle$ with the separation $\tau_{\mathrm{DD}}=N_t\sigma_t$. The stepwise (purple) and sinusoidal (blue) detunings $\Delta(t)$ are synchronized with the DD sequence. (c) Simulated infidelity $1-F_A^{(\sin)}$ for sinusoidal detuning with $N_t=12,16,20$ (blue diamonds, red circles, and yellow triangles). The purple curve shows $1-F_A^{(\mathrm{step})}$ for stepwise detuning with $N_t=10$. Parameters: $C=2$.
        }
        \label{fig_three_level}
        \vspace{-0.2cm}
    \end{figure}
    {\it Three-level register.---}
    Our protocol can be extended to a three-level $L$-type register where only the transition $\mathcal{T}_0$ is coupled to the cavity (Fig.~\ref{fig_three_level}(a)). $\mathcal{T}_1$ is far detuned and leads an empty cavity-like reflection $r_{\mathrm{off}}(\omega)=1-2\kappa/(\kappa-2i\omega)$, which induces a $\pi$ phase to the monochromatic pulse~\cite{Omlor2025, Nemoto2014}. Thus, the two amplitudes in a single round are always $|r_0(\omega)|\neq|r_1(\omega)|$. To achieve the entangling condition, as illustrated in Fig.~\ref{fig_simulation}(b), we apply a dynamical decoupling sequence of $2N$ $\pi$ pulses (green blocks) that periodically flip $|0\rangle$ and $|1\rangle$, with a pulse separation $\tau_{\mathrm{DD}}=N_t\sigma_t$. Meanwhile, a stepwise detuning $\Delta(t)$ (purple line), which switches between $+\Delta$ and $-\Delta$ synchronously with the DD timing, is applied. As a result, the effective reflections $\tilde{r}_0(\omega)=r_0(\omega)r_{\mathrm{off}}(\omega)$ and $\tilde{r}_1(\omega)=r_{\mathrm{off}}(\omega)r_1(\omega)$ after every two rounds of spin photon interaction, thereby requiring a different condition $\tilde{r}_+(\omega)=0$. For improved experimental compatibility, the stepwise detuning can be replaced by a smooth sinusoidal modulation $\Delta(t)=\tilde{\Delta}\sin(\nu t)$ (blue line) with $\nu=\pi/\tau_{\mathrm{DD}}$. Fig.~\ref{fig_three_level}(c) shows that the simulated fidelities $F_A^{(\sin)}$ under sinusoidal detuning with $N_t=12,\,16,\,20$ (blue diamonds, red circles, and yellow triangles) exceed $0.99$, while remaining bounded by the stepwise results $F_A^{(\mathrm{step})}$ with $N_t=10$ (see~\ref{Sec_SI_three}).
    
    {\it Experimental feasibility.---}
    The four-level system containing two transitions is a popular platform to implement spin-photon conditional gate~\cite{Dimer2007, Li2012}, demonstrating in atoms, silicon-vacancy centers in diamond~\cite{Bersin2024}, quantum dot system(QDs)~\cite{Warburton2013}, as well as molecular spin~\cite{GaitaArino2019}. 
    As an example, a self-assembled quantum dot operated in the Voigt geometry provides a four-level structure, where the two transitions with electronic ground states $|\uparrow\rangle \leftrightarrow |\uparrow\downarrow,\Uparrow\rangle$ and $|\downarrow\rangle \leftrightarrow |\uparrow\downarrow,\Downarrow\rangle$ have identical dipole moments~\cite{Warburton2013}. The corresponding transition frequencies can be continuously tuned via the in-plane magnetic field. In the bad-cavity regime of a Fabry-P\'erot cavity, this system can reach a Purcell factor $4g^2/(\kappa\gamma_{\mathrm{se}})\sim 10$~\cite{Antoniadis2023}, enabling efficient entangling operations even in the presence of pure dephasing $\gamma_{\phi}$, with an effective cooperativity $C>2$ achieved~\cite{Sun2016, Wells2019}. A FP cavity with $\eta_i\sim 0.97$ has also been reported~\cite{Tomm2021}. Typically, the transition frequency can be tuned to the telecom band~\cite{Yu2023, Laccotripes2024, Huang2025_op, Barbiero2024}, which significantly reduces the loss of the device.
    For the three-level spin, one candidate is the nitrogen-vacancy (NV) center in diamond~\cite{Doherty2013, Wu2016}, where only the transition $|0\rangle \leftrightarrow |E_{x/y}\rangle$ is tuned to selectively couple with the cavity~\cite{Nemoto2014, Omlor2025} with $C\sim 1$ having been achieved~\cite{Riedel2017, Yurgens2024}. The required $\Delta(t)$ can be implemented via a DC Stark shift by applying an oscillating electric field at frequency $\nu$, provided $\nu$ is far below the THz-scale zero-phonon line. In addition, some molecular defects share an NV-like ground-state triplet manifold and optical transitions~\cite{Bayliss2020, Kopp2024, Roggors2025}, while chemical synthesis enables more flexible level-structure engineering to realize effective four- or three-level schemes. 
    
    The overall encoding time is ultimately limited by the spin coherence. For the four-level scheme, the phase-encoding duration should remain well below $T_2^*$; the latter can be substantially extended by suppressing hyperfine coupling to the nuclear bath in both QDs and NV centers~\cite{Nguyen2023, Herbschleb2019}. The DD pulses employed in the three-level case suppress slow dephasing noise and thereby relax the constraint from $T_2^*$ to $T_2$, and the same strategy can also be extended to the four-level scheme.    
    Moreover, our protocol extends conventional one-shot architectures, differing by an external cycle that increases the effective optical path length. The more stringent phase-stability requirement could be met using established kilometer-scale techniques such as active stabilization with co-propagating referencing and common-path engineering, thereby ensuring a high-fidelity~\cite{Nardelli2025, LiuJL2024, Stolk2024}.
    
    {\it Conclusion and discussion.---}
    We propose a high-probability remote entanglement protocol to connect registers in the MCR based on repeated reflection-based spin-photon phase encoding. By recycling the reflected photon for multiple interactions with a detuned spin, a single spin imprints a conditional $\pi$ phase shift on the photon with high probability without requiring strong cooperativity. As the number of repetitions increases, both the success probability and the entanglement fidelity can be significantly enhanced in an ideal, lossless setup. Under our estimates of losses and imperfections in state-of-the-art setups, the protocol can be optimized to make the success probability remain acceptable and, in principle, approach unity as device performance continues to improve. This protocol is well-suited for applications from linking multi-spin registers~\cite{vanDam2017, Ruskuc2025}, where local entangling and re-initialization are slower~\cite{Liu2024}, to long-distance links where the propagation time dominates. Moreover, optimization favors width-scaling pulses with shorter duration, which imprint the spin-dependent response more efficiently and boost encoding rates even at moderate cooperativity. 
    This flexible tuning between fidelity and entanglement efficiency makes the protocol broadly applicable to solid-state spin platforms and provides a route toward photon-loss-tolerant distributed architectures based on moderate-cooperativity registers.
    
    {\it Acknowledgments.---}
    We acknowledge Carlos Munuera-Javaloy for helpful discussions. Y. L. thanks Qingyun Cao for useful discussions on the experimental aspects of NV centers. This work is supported by the BMBF under the funding program ‘quantum technologies—from basic research to market’ in the project Spinning (project No.~13N16215) and project CoGeQ (project No.~13N16101).  This work was supported by EU-project C-QuENS (Grant No.~101135359) and EU-Project SPINUS (Grant No.~101135699). The authors acknowledge support by the state of Baden-W\"{u}rttemberg through bwHPC and the German Research Foundation (DFG) through grant no INST 40/575-1 FUGG (JUSTUS 2 cluster)

    \newpage
    \bibliography{reference}
    \onecolumngrid
    \newpage
    \appendix
    \begin{center}
        {\Large\bfseries Supplementary Materials}
    \end{center}
    \vspace{3ex}
    \twocolumngrid

    \section{Phase-encoding based heralded entangling}
    \label{Sec_SI_phased_encoding}
    In this section, we study a spin-cavity register in which a single-sided cavity hosts a mode $\hat{c}$ that interacts with a four-level spin system, as sketched in Fig.~1(a). The relevant register $R_M$ (M=A,~B) consists of two transitions $\mathcal{T}_{s,M}$ ($s\in\{0,1\}$), each connecting a ground state $|s\rangle_M$ to an excited state $|e_s\rangle_M$. Throughout, we assume the two transitions are symmetric, with identical dipole moments, so that they share the same cavity coupling strength and the same excited-state decay rate. Meanwhile, the stationary qubit is encoded in the ground-state manifold $\{|0\rangle_M,|1\rangle_M\}$, and each register $R_M$ is prepared in the equal superposition $|+\rangle_M=(|0\rangle_M+|1\rangle_M)/\sqrt{2}$. Using two such registers, we implement a phase-encoding, heralded entanglement scheme: a Mach-Zehnder interferometer coherently routes an input photon through the two spin-cavity registers $R_A$ and $R_B$, and a detection click at the output ports heralds remote entanglement between the two stationary qubits.
    
    In the long-pulse limit, a single incident photon $|u\rangle_A= a^\dagger_A|\mathrm{vac}\rangle $ is sent from port A, and then we could write the initial state as
    \begin{align}
        \frac{a^\dagger_A}{2} (|00\rangle+|01\rangle+|10\rangle+|11\rangle),
    \end{align}
    with $|mn\rangle = |m\rangle_A|n\rangle_B$. The photon is split at the Beam splitter, and the BS transforms the state to 
    \begin{align}
        \frac{ia^\dagger_A+a^\dagger_B}{2\sqrt{2}} (|00\rangle+|01\rangle+|10\rangle+|11\rangle),
    \end{align}
    due to the transforms ${a}_A^\dagger\rightarrow({a}_B^\dagger+i{a}_A^\dagger)/\sqrt{2}$ and ${a}_B^\dagger\rightarrow(i{a}_B^\dagger+\hat{a}_A^\dagger)/\sqrt{2}$.
    And then, the photon will be sent to the registers, and the following conditional reflection-based encoding is described as
    \begin{align}
        a_M^\dagger|0\rangle_M \Rightarrow r_0 a_M^\dagger|0\rangle_M,\quad
        a_M^\dagger|1\rangle_M \Rightarrow r_1 a_M^\dagger|1\rangle_M.
    \end{align}
    Here, $r_s$ is a complex number indicating the modification implemented to the photon mediated by the spin-cavity register, and $|r_s|^2<1$ is the probability of the photon being reflected. Now we can express that the state is 
    \begin{align}
        &\frac{ia^\dagger_A} {2\sqrt{2}}\left(r_0|00\rangle+r_0|01\rangle+r_1|10\rangle+r_1|11\rangle\right)\nonumber\\
        +& \frac{a^\dagger_B} {2\sqrt{2}}\left(r_0|00\rangle+r_1|01\rangle+r_0|10\rangle+r_1|11\rangle\right).
        \label{eq_SI_state}
    \end{align}
    Then the photon is guided back and interferes at the BS, using the inverse transforms $a_A^\dagger \rightarrow(-ia_A^\dagger+a_B^\dagger)/\sqrt{2}$ and $a_B^\dagger \rightarrow(a_A^\dagger-ia_B^\dagger)/\sqrt{2}$, resulting in a state
    \begin{align}
        &\frac{a_A^\dagger+ia_B^\dagger} {4}\left(r_0|00\rangle+r_0|01\rangle+r_1|10\rangle+r_1|11\rangle\right)\nonumber\\
        +& \frac{a_A^\dagger-ia_B^\dagger} {4}\left(r_0|00\rangle+r_1|01\rangle+r_0|10\rangle+r_1|11\rangle\right).
    \end{align} 
    And we could get the final state as
    \begin{align}
        |\psi_f\rangle = \frac{a^\dagger_A}{\sqrt{2}}\left(r_+|\Phi^+\rangle+r_-|\Phi^-\rangle + r_+|\Psi^+\rangle\right)+\frac{ia^\dagger_B}{\sqrt{2}}r_-|\Psi^-\rangle.
        \label{eq_SI_psi_f}
    \end{align}
    which is exactly the results in the main text.
    With the Bell states are $|\Phi^{\pm}\rangle=(|00\rangle\pm|11\rangle)/\sqrt{2}$ and $|\Psi^{\pm}\rangle=(|01\rangle\pm|10\rangle)/\sqrt{2}$. 
    Under the entangling condition, the probability of one photon being detected at each port is $P_A=P_B$, giving a total success probability of 
    \begin{align}
        P_t = P_A+P_B = |r_-|^2 = \frac{1}{4}|r_0-r_1|^2 = |r_0|^2=|r_1|^2.
        \label{eq_SI_P_a_B}
    \end{align}
    In the $N$-round interaction-based protocol, the success probability is $P_t=|r_0|^{2N}$.
    This means that the reflected photons are converted with unit probability to useful photons that are then detected to herald the entangled state. This is the reason that the proposal operates at the large detuning regime, where the decoherence relevant to the excited state are strongly suppressed to maximized the reflections.
    
    Considering an input envelope in the frequency domain, the single input photon is
    \begin{align}
        |u\rangle_A=\int d\omega\tilde{u}(\omega) a^\dagger_A(\omega)|\mathrm{vac}\rangle
        =\int dt {u}(t) a^\dagger_A(t)|\mathrm{vac}\rangle,
    \end{align} 
    with the Fourier transform
    \begin{align}
        {u}(t) = \frac{1}{\sqrt{2\pi}} \int d\omega \tilde{u}(\omega) e^{-i\omega t}.
    \end{align}
    Then the conditional reflections in both registers are
    \begin{align}
        &\int d\omega \tilde{u}(\omega)a_M^\dagger(\omega)|0\rangle_M \Rightarrow \int d\omega r_0(\omega) \tilde{u}(\omega)a_M^\dagger(\omega)|0\rangle_M,\nonumber\\
        &\int d\omega \tilde{u}(\omega)a_M^\dagger(\omega)|1\rangle_M \Rightarrow \int d\omega r_1(\omega)\tilde{u}(\omega)a_M^\dagger(\omega)|1\rangle_M,
        \label{eq_SI_con_ref_w}
    \end{align}
    resulting in a final state
    \begin{widetext}
        \begin{align}
            |\psi_f\rangle = \frac{1}{\sqrt{2}}\int d\omega \Big\{\left[|\Phi^+\rangle r_+(\omega)+|\Phi^-\rangle r_-(\omega) + |\Psi^+\rangle r_+(\omega)\right]\tilde{u}(\omega)a^\dagger_A(\omega)+i|\Psi^-\rangle r_-(\omega) \tilde{u}(\omega)a^\dagger_B(\omega)\Big\}|\mathrm{vac}\rangle.
            \label{eq_SI_pis_f_iden}
        \end{align}
    \end{widetext}
    
    Thus, the click probabilities at each port are
    \begin{align}
        P_M &= \int d\omega'\langle \psi_f|a^\dagger_M(\omega')a_M(\omega')|\psi_f\rangle\nonumber\\
        &= \int d\omega'|\langle 1_M(\omega')|\psi_f\rangle|^2,
    \end{align}
    with $|1_M(\omega')\rangle = a_M^\dagger(\omega')|\mathrm{vac}\rangle$, and it is valid that there is at most one photon is detected at each port.
    \begin{align}
        P_A &= \frac{1}{2}\int d\omega \left[|r_-(\omega)|^2+2|r_+(\omega)|^2\right]|\tilde{u}(\omega)|^2 \nonumber\\
        P_B &= \frac{1}{2}\int d\omega |r_-(\omega)|^2|\tilde{u}(\omega)|^2\, .
    \end{align}
    And the corresponding states of joint state with one click on each port are given by 
    \begin{align}
        \frac{1}{P_M} \int d\omega' a_M(\omega') |\psi_f\rangle\langle\psi_f|a_M^\dagger(\omega')\, .
    \end{align}
    Tracing out the degree of photon, one can obtain the heralded two-register state as
    \begin{align}
        \rho_A&= \frac{1}{2P_A}\int d\omega'|\tilde{u}(\omega')|^2|\psi_A(\omega')\rangle\langle \psi_A(\omega')|\\
        \rho_B & = |\Psi^-\rangle\langle\Psi^-|
    \end{align}
    with state $|\psi_A(\omega')\rangle=r_+(\omega')(|\Phi^+\rangle+|\Psi^+\rangle)+r_-(\omega')|\Phi^-\rangle$. Thus, one click at port B heralds a perfect Bell state ($F_B=1$), while at port A, one gets a mixed state.
    Approaching the entangling condition $r_+(\omega)=0$~\cite{Omlor2025}, $\rho_A$ is approaching $|\Phi^-\rangle\langle\Phi^-|$, with the corresponding fidelity is
    \begin{align}
        F_A = \frac{\int d\omega |r_-(\omega)|^2|\tilde{u}(\omega)|^2}{2P_A} = \frac{P_B}{P_A}.
    \end{align}
    The results with the long pulse limit can be got by setting $\tilde{u}(\omega) = \delta(\omega)$.
    
    \section{The input-output relation}
    \label{Sec_SI_input_out}
    In this section, we discuss the reflection of a two-level system coupled with the cavity. A Jaynes-Cummings model Hamiltonian is 
    \begin{align}
        H = \omega_a |e\rangle \langle e| + \omega_c c^\dagger c + g(c^\dagger \sigma_-+c\sigma_+)\, ,
    \end{align}
    with $\sigma_- = (\sigma_+)^\dagger=|g\rangle \langle e|$.
    Under the single-excitation subspace (at most one photon exists in the system), the interaction Hamiltonian to describe a single incident photon (incident photon) sent from the left side of the cavity is 
    \begin{align}
        H_{\mathrm{int}} =  i\sqrt{\kappa_l} \left( \alpha_{\mathrm{li}}^*(t) c - \alpha_{\mathrm{li}}(t) c^\dagger \right).
    \end{align}
    Here, the input envelope from the left side is regarded as a weak classical driving~\cite{Borges2016}
    \begin{align}
        \alpha_{\mathrm{li}}(t) = u(t) e^{-i\omega_p t}\, ,
    \end{align}
    with $\omega_p$ the central frequency of the incident photon and $u(t)$ is the envelope of the input photon. Moving into the interaction picture with respect to $H_0 = \omega_p |e\rangle \langle e| + \omega_p c^\dagger c$, we get the effective Hamiltonian 
    \begin{align}
        H_I =& \Delta |e\rangle \langle e| + \Delta_c c^\dagger c + g(c^\dagger \sigma_-+c\sigma_+) \nonumber\\
        &+ i\sqrt{\kappa_l} \left( u^*(t)c - u(t) c^\dagger \right)
    \end{align}
    with the detuning $\Delta_{c}=\omega_{c}-\omega_p$ and $\Delta=\omega_a-\omega_p$.
    Typically, we include additional non-Hermitian terms $-i(\gamma_{\mathrm{se}}/2)|e\rangle\langle e|$, $-i\gamma_\phi|e\rangle\langle e|$, and $-i\kappa c^\dagger c$, which account for spontaneous emission, pure dephasing of the two-level system, respectively. The total cavity-field loss rate $\kappa=\kappa_r+\kappa_l+\kappa_i$.
    Here, the input incident photon is regarded as a weak drive, the evolution of the single-excitation state~\cite{Borges2016}
    \begin{align}
        |\Psi(t)\rangle &= c_{g1}(t) |g,1_c\rangle + c_{e0}(t) |e,0_c\rangle
    \end{align}
    is
    \begin{align}
        \dot{c}_{g1} &= -i\left(\Delta_c - i\frac{\kappa}{2}\right) c_{g1} -i g c_{e0} - \sqrt{\kappa_l} u(t)\, ,  \\    
        \dot{c}_{e0} &= -i\left(\Delta - i\frac{\gamma}{2}\right) c_{e0} -i g c_{g1}.
    \end{align}
    The initial conditions of the amplitudes are $c_{g1} = c_{e0} =0$, and $\gamma = \gamma_{\mathrm{se}}+2\gamma_{\phi}$. To solve the evolution, we employ the input-output relations at both sides of the cavity~\cite {Agarwal2013, Collett1984, Gardiner1985, Reiserer2015}
    \begin{align}
        u_{\mathrm{ro}}(t) &= \sqrt{\kappa_r} c_{g1}(t),\\
        u_{\mathrm{lo}}(t) &= u(t) + \sqrt{\kappa_l} c_{g1}(t).
    \end{align}
    
    Moving to the frequency domain, the Fourier transform of the envelope is
    \begin{align}
        \tilde{u}(\omega) = \frac{1}{\sqrt{2\pi}}\int dt  u(t)e^{-i\omega t}
    \end{align}
    where the frequency $\omega$ is detuning of the incident photon between $\omega_p$.
    Transforming each equation to the frequency domain gives
    \begin{align}
        &\left(\Delta_c-\omega-i\frac{\kappa}{2}\right) \tilde{c}_{g1}(\omega) + g\tilde{c}_{e0}(\omega) -i\sqrt{\kappa_l} \tilde{u}(\omega) =0\, ,\nonumber\\
        &\left(\Delta-\omega-i\frac{\gamma}{2}\right) \tilde{c}_{e0}(\omega) + g\tilde{c}_{g1}(\omega) =0.
    \end{align}
    We could solve the relevant reflection and transmission as
    \begin{align}
        r(\omega) & = \frac{\tilde{u}_{\mathrm{lo}}(\omega)}{\tilde{u}(\omega)} \nonumber\\
        &= 1- \frac{2\kappa_l \left[2i(\Delta - \omega) + \gamma\right]}{[2i(\Delta_c-\omega)+\kappa][2i(\Delta-\omega)+\gamma]+4g^2}\nonumber\\
        t(\omega) & = \frac{\tilde{u}_{\mathrm{ro}}(\omega)}{\tilde{u}(\omega)} \nonumber\\
        &= \frac{-2\sqrt{\kappa_l \kappa_r} \left[i2(\Delta - \omega) + \gamma\right]}{[2i(\Delta_c-\omega)+\kappa][2i(\Delta-\omega)+\gamma]+4g^2}.
        \label{Eq_SI_rt}
    \end{align}
    This gives Eq.(1) in the main text by setting the central frequency of the incident photon to be resonant with the cavity $\omega_c=\omega_p(\Delta_c=0)$. 
    
    \subsection{Suppression of the excitation by large detuning}
    Meanwhile, we could get the amplitude of the excited state in the frequency domain as 
    \begin{align}
        c_{e0}(\omega) = \frac{4ig\sqrt{\kappa_l}}{[2i(\Delta_c-\omega)+\kappa][2i(\Delta-\omega)+\gamma]+4g^2}\tilde{\alpha}(\omega).
    \end{align}
    For simplicity, in the long-pulse limitation, the frequency domain transfer function from the input envelope to the excited state amplitude is
    \begin{align}
        c_{e0} = \frac{4 i g \sqrt{\kappa_l}}{\kappa\left(\gamma+2 i \Delta\right)+4 g^2}\alpha_0.
    \end{align}
    The steady state population is $p_e^{\mathrm{ss}}=|c_{e0}^{\mathrm{ss}}|^2$ which evaluates to
    \begin{align}
        p_e^{\mathrm{ss}}(\Delta)=\frac{16 g^2 \kappa_l}{\left(\kappa \gamma+4 g^2\right)^2+\left(2 \kappa \Delta\right)^2} |\alpha_0|^2\, .
    \end{align}
    This is a Lorentzian in $\Delta$ centered at zero, which decreases as $|\Delta|$ increases.
    The far-detuned asymptotic scaling follows directly
    \begin{align}
        p_e^{\mathrm{ss}}(\Delta)\sim \frac{4 g^2 \kappa_l}{\kappa^2 \Delta^2} |\alpha_0|^2 \quad \mathrm{for } |\Delta|\to \infty\, .
    \end{align}
    The result shows that, under cavity resonant driving, the steady-state excited-state population is maximized near $\Delta=0$ and decreases as $1/\Delta^2$ for large detuning.
    
    \section{Reflection-based phase encoding}
    \label{Sec_SI_reflcetion_based}
    \subsection{Probability achieved}
    \label{Subsec_SI_probability}
    In our proposal, the reflected photons serve as the resources for the entangling operation in a single-sided cavity ($\kappa_l=\kappa$). Under the long-pulse (monochromatic) condition with input $\alpha_0 e^{-i\omega_p t}$, the spin-state-dependent response is characterized by the reflection coefficient.
    \begin{align}
        r_s & = 1-\frac{2\kappa(2i\Delta_s + \gamma)}{\kappa(2i\Delta_s + \gamma)+4g^2}\, .
    \end{align}
    Choosing a symmetric case, $\Delta = \Delta_0=-\Delta_1$, and we get $r_0=r_1^*$
    with $O = 2 \Delta/\gamma$. Under the large detuning limit $O\gg1$, we get 
    \begin{align}
        r_0 &= -1 -i \frac{2C}{O} +\frac{2C(1+C)}{O^2} +\mathcal{O}\left(\frac{1}{O^3}\right) \nonumber\\
        &\approx - Re^{i\theta},\quad (R
        >0).
    \end{align}
    with $\theta\approx2C/O= C\gamma/\Delta$ is the phase picked in each round.
    To satisfy the entangling condition $(r_0)^{N}+(r_1)^{N}=0$, the number of interaction rounds is determined as
    \begin{align}
        N = \frac{\pi}{2\theta} = \frac{\pi\Delta}{2C\gamma}\label{eq_SI_N}.
    \end{align}
    Now we can see that the reflection efficiency in a single round is
    \begin{align}
        R^2 = 1 - \frac{4C}{O^2} + \mathcal{O}\left(\frac{1}{O^3}\right)
        \label{eq_SI_R}
    \end{align}
    approaches 1 when $\Delta/\gamma\rightarrow\infty$ increases, and requiring $N\rightarrow\infty$. 
    Substituting Eq.~\eqref{eq_SI_N} into Eq.~\eqref{eq_SI_R} yields
    \begin{align}
        R^2 = 1 - \frac{\pi^2}{4C} \frac{1}{N^2} + \mathcal{O}\left(\frac{1}{N^3}\right).
    \end{align}
    Thus, we get the total reflection probability is
    \begin{align}
        P_t^{(\delta)}=R^{2N}= \left(1 - \frac{\pi^2}{4C} \frac{1}{N^2} + \mathcal{O}\left(\frac{1}{N^3}\right)\right)^N.
    \end{align}
    which shows $P_t^{(\delta)} \to 1$ as $N\to\infty$. This is proved as setting $A=\pi^2/(4C)$, one has $\ln R^{2N}=N\ln(1-A/N^2+\mathcal{O}(1/N^3))=-A/N+\mathcal{O}(1/N^2)\to0$. Therefore, the leading term $R^{2N}\approx\exp(-A/N)$ suggests that $P_t^{(\delta)}$ eventually increases with $N$ and approaches unity. And the probability $P_t$ for the spectral width pulse will also gain a high probability.

    Note that this entanglement success probability, which approaches unity, is particularly attractive for systems exhibiting moderate cooperativity ($C\sim1$). However, achieving this objective necessitates perfect and ideal experimental conditions, namely an infinite number of rounds of interactions between a single incident photon and the spin register. Nevertheless, this offers a feasible approach for achieving high-success-rate entanglement in moderate-coherence systems, representing a step toward fault-tolerant quantum computing within the moderate cooperativity spin systems. 
    
    \subsection{Repeated spin-photon encoding scheme}
    In the repeated scheme of spin-photon encoding, the state-dependent conditional encoding (Eq.~\ref{eq_SI_con_ref_w}) after the $n$-th interaction is
    \begin{align}
        &\int d\omega \tilde{u}(\omega)a_M^\dagger(\omega)|0\rangle_M \Rightarrow \int d\omega r_0(\omega)^n \tilde{u}(\omega)a_M^\dagger(\omega)|0\rangle_M,\nonumber\\
        &\int d\omega \tilde{u}(\omega)a_M^\dagger(\omega)|1\rangle_M \Rightarrow \int d\omega r_1(\omega)^n\tilde{u}(\omega)a_M^\dagger(\omega)|1\rangle_M\, .
    \end{align}
    In this case, we have to define a new parameter
    \begin{align}
        r_{\pm}(\omega) = \frac{1}{2}[r_0(\omega)^n\pm r_1(\omega)^n].
    \end{align}
    
\begin{figure}[t]
        \centering
        \includegraphics[width=0.45\textwidth]{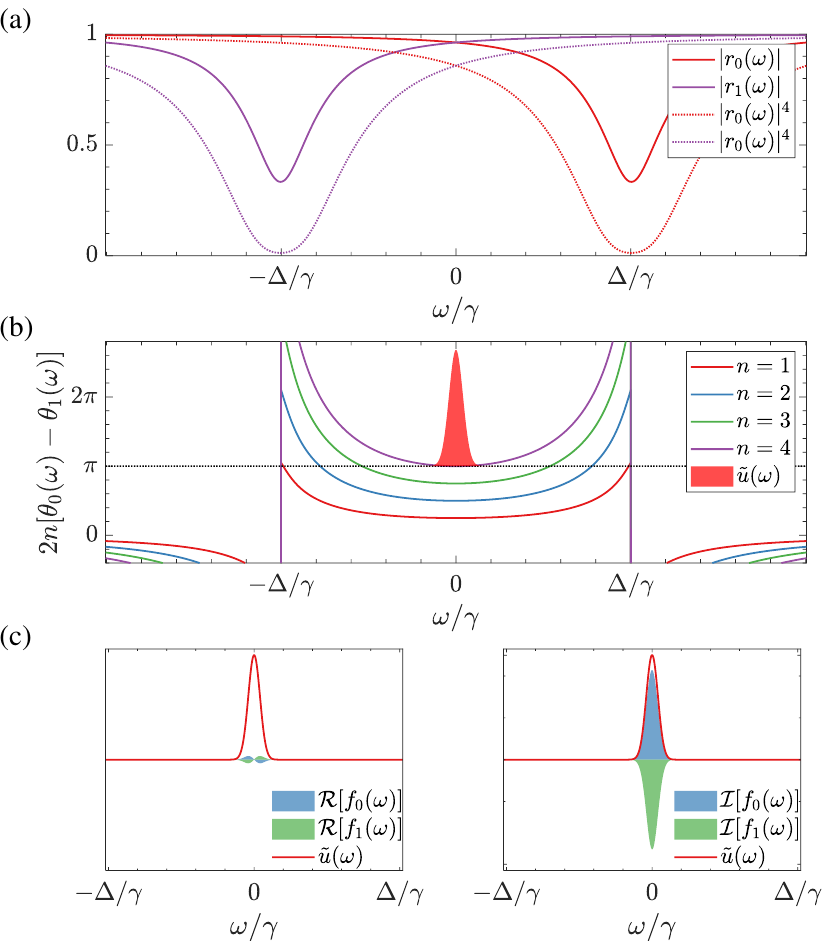}
        \caption{Repeated spin-photon encoding process, providing additional details that supplement Fig.~1. (a) The reflection amplitude $|r_s(\omega)|$ (solid line) and $|r_s(\omega)|^4$ (dotted line) with $s=0$(red) and $s=1$(purple). (b) The relative phase $2n[\theta_0(\omega)-\theta_1(\omega)]$ accumulated after the $n$th interaction round, reaching $\pi$ at $N=4$. (c) The real (left) and imaginary (right) parts of the final envelope $f_s(\omega)$ in the frequency domain. The solid red line indicates the input envelope. Parameters: $\kappa=200\gamma$, $C=2$, $\sigma_\omega=\gamma/5$, and $\Delta/\gamma = 4.9873$.}
        \label{fig_SI_repeated_scheme}
    \end{figure}

    Now, fulfilling the entangling condition requires choosing an optimal repetition number $n(=N)$ and a frequency window in which $r_+(\omega)=0$ is approximately satisfied. Fig.~\ref{fig_SI_repeated_scheme}(a) shows the state-dependent reflection amplitudes after one round, $|r_0(\omega)|$ and $|r_1(\omega)|$, and after four rounds, $|r_0(\omega)|^4$ and $|r_1(\omega)|^4$. The two spectra coincide at $\omega=0$. The minima of $|r_0(\omega)|^2$ and $|r_1(\omega)|^2$ occur near $\omega=\pm\Delta$, where $\Im[r_s(\omega)]=0$. The phases of $r_0(\omega)$ and $r_1(\omega)$ differ by $2\theta(\omega)$, and this relative phase accumulates over repeated photon-register interactions. When the total phase reaches $\pi$, the entangling condition is met (Fig.~\ref{fig_SI_repeated_scheme}(b)). Figure~\ref{fig_SI_repeated_scheme}(c) plots the frequency-domain envelopes of the final states $f_s(\omega)=[r_s(\omega)]^n\tilde{u}(\omega)$, for which the phase encoding is nearly achieved when $f_0(\omega)+f_1(\omega)\approx0$. Parameters: $\kappa=200\gamma$, $C=2$, $\sigma_\omega=\gamma/5$, $\Delta/\gamma=4.987$, and $n=4$.

    \subsection{The repetitions}
    In this section, we will discuss the repetitions allowed for different cooperativities. For monochromatic pulse $(\omega=0)$, the reflection amplitude is
    \begin{align}
        r(\Delta) &= \frac{C-1-i\delta}{C+1+i\delta},
    \end{align}
    with $\delta=2\Delta/\gamma$.
    Using the entangling condition, $r(\Delta)^N+r(-\Delta)^N=0$ gives the phase $\arg[-r(\Delta)] = \pi/2N$, making 
    \begin{align}
        A=\tan\left(\frac{\pi}{2N}\right) = -\frac{2\delta C}{C^2-\delta^2-1},
    \end{align}
    and 
    \begin{align}
        A\delta^2 -2\delta C -A(C^2-1) &= 0,
    \end{align}
    whose solutions of the detuning are
    \begin{align}
        \Delta_{\pm} &= \frac{\gamma}{2A}\left[C \pm \sqrt{C^2+\left(C^2-1\right)A^2}\right].
    \end{align}
    A real solution requires
    \begin{align}
        C^2+\left(C^2-1\right)A^2 &\ge 0
    \end{align}
    giving the condition 
    \begin{align}
        C^2\ge \frac{A^2}{1+A^2} = \sin^2\left(\frac{\pi}{2N}\right).
    \end{align}
    Thus, the cooperativity requirement can be written as $C \ge \sin(\pi/2N)$, which sets a lower bound on $C$ for a given repetition number $N$. For single-round entangling ($N=1$), one must have $C \ge 1$. Hence, in the low-cooperativity regime ($C<1$), the condition can only be met with multiple repetitions ($N>1$). This is consistent with the minimal starting repetitions $N=1,2,4$ for $C=2,1,0.5$ in Fig.~\ref{fig_simulation}(e), and accounts for the blank regions in Fig.~\ref{fig_SI_cavity_round_loss_II}(a,b).
    
    \section{The duration optimizing}
    \label{Sec_SI_duration}
    In this section, we are going to perform a few simulations about the limitations of the protocol. In our proposal, with an infinite repetition, the total success probability approaches 1. However, despite the losses we discussed before, the repetition of the interactions is also limited by the coherence time of the spin ground state. In the four-level system, the total duration of the repeated phase encoding should be much smaller than the dephasing time $T_2^*$. For the solid-state spin systems like quantum dots and NV centers, this inhomogeneous dephasing noise mainly comes from the hyperfine interaction with the nuclear bath and modify the energy gap of the ground state spin, resulting in a time-varying detuning $\Delta+\delta(t)$. Therefore, this bias $\delta(t)$ causes a fluctuation in the phase accumulated in each round. The overall conditional phase difference deviates from $\pi$ by a random amount. In the end, the entangling fidelity will decrease accordingly, while the success probability will remain nearly unchanged because the dephasing only alters the phase. 
    
    To operate the entangling within the constraints of the $T_2^*$, two methods should be used. The first point is to make the duration as short as possible. The total encoding duration is
    \begin{align}
        T_{\mathrm{pe}} = NN_t\sigma_t = \frac{NN_t}{\sigma_{\omega}},
        \label{Eq_SI_TE}
    \end{align}
    where $N$ is the repetition; $N_t\sigma_t$ is the length of the temporal envelop of Gaussian pulse; the width $\sigma_t = 1/\sigma_\omega$. Here we assume the absence of the pure dephasing $(\gamma_\phi=0)$ of the transition, and $\tau_e=1/\gamma$ is the lifetime of the excited state. 
    Here, we use a Gaussian-shaped pulse in the frequency domain
    \begin{align}
        \tilde{u}(\omega)=(\pi\sigma_\omega^2)^{-1/4}\exp(-\omega^2/2\sigma_\omega^2),
    \end{align}
    with $\tilde{u}(\omega)^2$ a zero-mean Gaussian distribution with standard deviation $\sigma_{\omega}$. And the temporal shape is
    \begin{align}
        u(t) = (\pi\sigma_t^2)^{-1/4}\exp\left(-t^2/2\sigma_t^2\right).
    \end{align}
    
    \begin{figure}[t]
        \centering
        \includegraphics[width=0.48\textwidth]{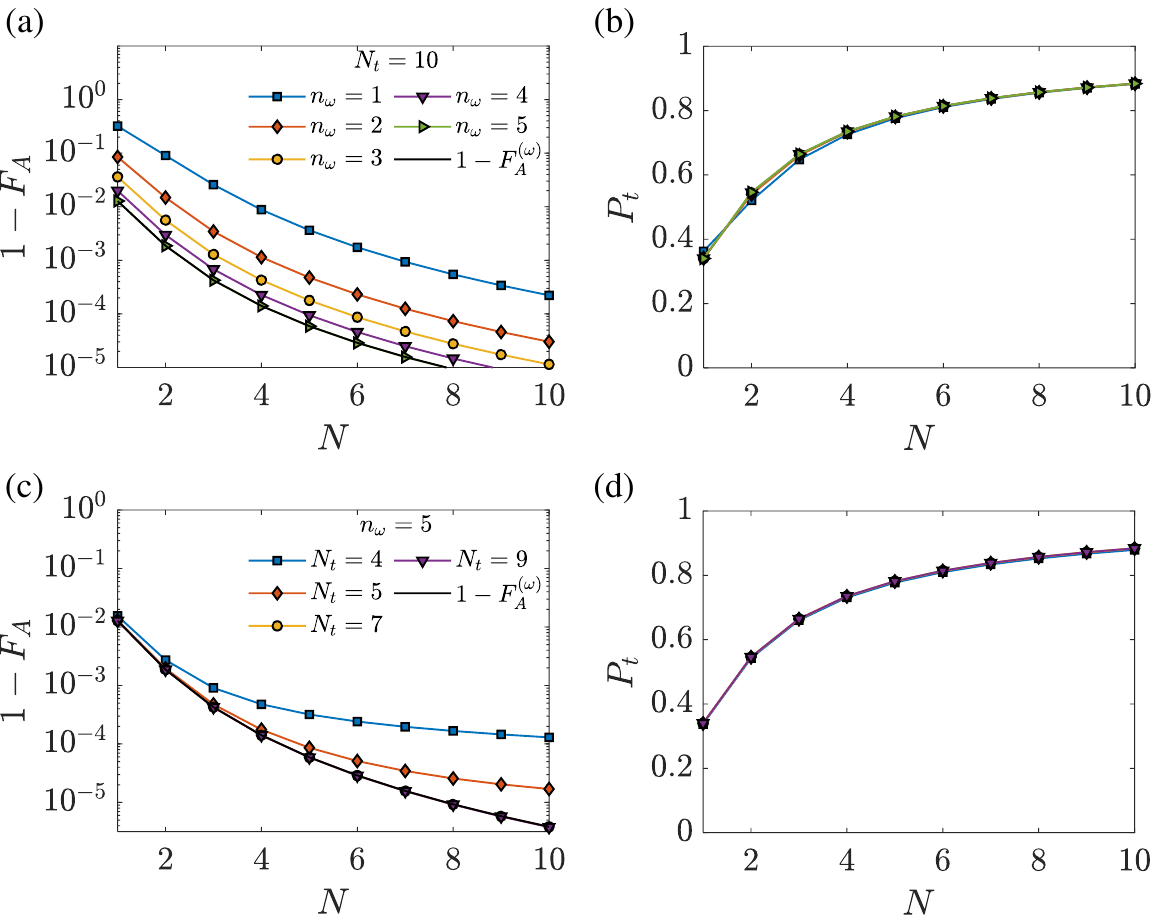}
        \caption{The simulated infidelity $1-F_A$(a) and probability(b) with the pulses cover different numbers of temporal widths and spectral width, i.e., $N_t$ and $n_\omega$}
        \label{fig_SI_duration}
    \end{figure}
    
    \subsection{Simulation of the duration}
    We first examine how the choice of $\sigma_\omega=\gamma/n_\omega$ influences the achievable fidelity. As shown in Fig.~\ref{fig_SI_duration}(a), increasing $n_\omega$ results in a pulse that is more spectrally localized within the entangling window, which in turn improves the fidelity. This improvement can be understood intuitively: a more spectrally confined pulse better satisfies the entangling condition by suppressing the spectral weight in the undesired channel, i.e., by reducing
    \begin{align}
        S=\int d\omega |r_+(\omega)|^2|\tilde{u}(\omega)|^2.
        \label{eq_SI_S}
    \end{align}
    Moreover, we present the simulated infidelity $1-F_A$ for different values of $N_t$, which corresponds to varying the pulse duration, as shown in Fig.~\ref{fig_SI_duration}(c). As $N_t$ increases, the overall simulation time window more fully encompasses the temporal length of a single pulse, leading to improved fidelity. It can be observed that for $N_t = 10$, the fidelity obtained from the time-domain simulation already converges to the reference result obtained in the frequency domain.
    
    \subsection{Increasing the phase encoding rate with width-scaling scheme}
\begin{figure}[t]
        \centering
        \includegraphics[width=0.48\textwidth]{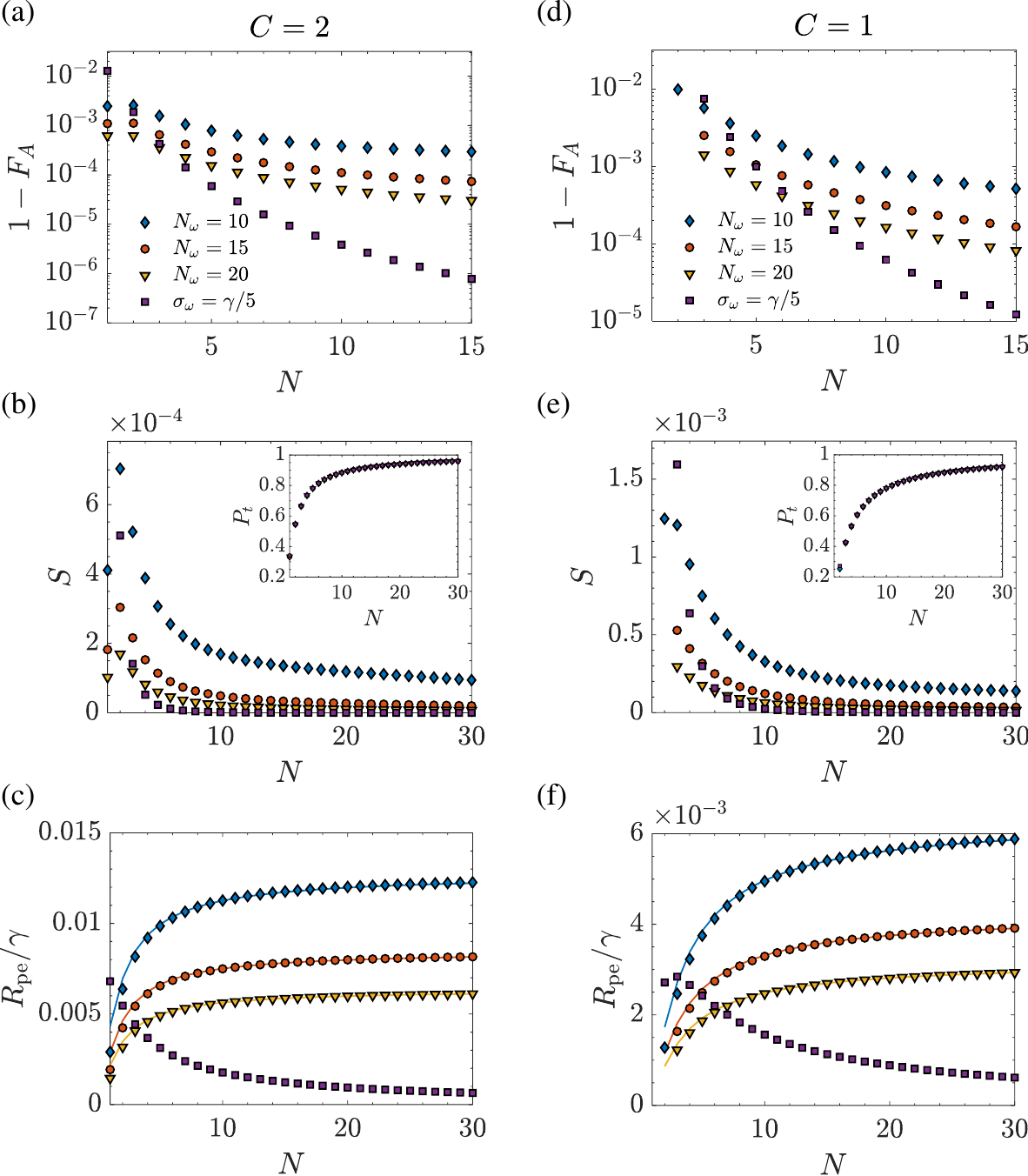}
        \caption{The simulated infidelity $1-F_A^{(t)}$(a) and probability(the inserts in (b)) with different spectral width $\sigma_\omega = \Delta/N_\omega$, with $N_\omega = 10, 15,  20$ and $\sigma_\omega = \gamma/5$ are presented by blue diamonds, orange circles, yellow triangles and purple squares. Here $C=2$. (b,c) show the corresponding metric of entangling condition $S$ and the phase encoding rates $R_{\mathrm{pe}}$. The results of $C=1$ are shown in (d-f).}
        \label{fig_SI_duration_II}
    \end{figure}

    The fidelity over the frequency domain is
    \begin{align}
        F_A = \frac{\int d\omega  |r_-(\omega)|^2|\tilde{u}(\omega)|^2}{\int d\omega \left[|2r_+(\omega)|^2+ |r_-(\omega)|^2\right]|\tilde{u}(\omega)|^2}.
    \end{align}
    And the value $S$~\eqref{eq_SI_S} indicates the overlap between the input pulse $\tilde{u}(\omega)$ and the entangling window. As we show in Fig.~\ref{fig_scheme}(a), the entangling window (the gray region) is localized around $\omega=0$, and bounded by the detuning $\Delta$, which is obtained by setting $\Im{r_{0/1}(\omega)}=0$. When the optimal detuning $\Delta=2NC\gamma/\pi$ scales linearly with the repetition number $N$, the entangling window broadens proportionally. This linear broadening is matched by choosing a pulse bandwidth that scales in the same way, $\sigma_{\omega}=\Delta/N_\omega$. As a result, the spectral-overlap metric $S$ remains nearly unchanged as $N$ increases(see Fig.~\ref{fig_SI_duration_II}(b,e)), yielding an almost $N$-independent fidelity, as shown in Fig.~\ref{fig_SI_duration_II}(a,d). With this width-scaling scheme, the temporal width $\sigma_t=1/\sigma_\omega$ decreases proportionally with $1/\Delta$, and the total phase-encoding duration $T_{\mathrm{pe}}=N N_t\sigma_t$ becomes independent of $N$. One then obtains the fixed encoding duration
    \begin{align}
        T_{\mathrm{pe}}^{N_\omega}=\frac{N_tN_\omega\pi}{2C\gamma},
    \end{align}
    which means that the encoding duration will be fixed. In contrast, the encoding using a fixed spectral width takes a duration
    \begin{align}
        T_{\mathrm{pe}}^{n_\omega} = \frac{NN_tn_\omega}{\gamma},
    \end{align}
    which is linear with $N$. And the success probabilities over two schemes are almost the same(see inserts in~\ref{fig_SI_duration_II}(b,e)). And the corresponding phase encoding rates $R_{\mathrm{pe}} = P_t/T_{\mathrm{pe}}$ are given in~Fig.~\ref{fig_SI_duration_II}(c,f), in which the width-scaling scheme exhibits an increasing rate due to the fixed duration and increasing probability. While the width-fixed scheme behaves totally opposite, and at a very low level. Typically, the ratio between them is
    \begin{align}
        \frac{R_{\mathrm{pe}}^{N_\omega}}{R_{\mathrm{pe}}^{n_\omega}} = \frac{n_\omega}{N_\omega} \frac{2CN}{\pi},
        \label{eq_SI_R_ratio}
    \end{align}
    which means that the width-fixed scheme exhibits a larger efficiency of phase encoding with $N$ and $C$ increases. As we show in Fig.~\ref{fig_SI_duration_II}(c,f), the solid lines are the results given by~\eqref{eq_SI_R_ratio}, which matches the numerical results well.
    
    Overall, the width-scaling strategy trades an acceptable reduction in fidelity for achieving a repeated spin-photon phase-encoding within a fixed total duration. In particular, by letting the pulse bandwidth scale with the optimal detuning, the temporal width of each incident photon can be reduced as the repetition number increases, which prevents the total phase-encoding duration from growing with the number of rounds. The achievable total duration is determined by several experimental and protocol parameters, including the cooperativity $C$, the chosen scaling parameter $N_\omega$, and the number of temporal widths $N_t$. 
    To realize a short overall phase-encoding duration, one direct strategy is to operate at larger cooperativity and choose a smaller $N_\omega$. As an explicit example, for a quantum-dot system with $C=2$, choosing $N_t=10$ and $N_\omega=10$ yields a total duration $T=40\tau_e$ for the phase-encoding. With $\tau_e=1/\gamma \sim 1$~ns, this corresponds to a total phase-encoding time of order $40$~ns, which can satisfy the practical requirement $T_{\mathrm{pe}} \ll T_2^*$. Since the cooling or narrowing of the nuclear-spin bath can substantially extend $T_2^*$, and values exceeding $600$~ns have been reported~\cite{Nguyen2023}, with further improvements expected as experimental techniques advance. 
    
    For three-level register implementations, one potential platform is the NV center in a diamond. In this case, the dominant low-frequency dephasing noise originates from the hyperfine interaction with the nuclear spin bath, and can be strongly suppressed by the applied dynamical-decoupling pulse sequences during the protocol. Specifically, considering a static detuning perturbation such that the effective detuning during an interaction is shifted from $\Delta$ to $\Delta+\delta$, where $\delta$ is approximately constant over two successive encoding segments. The total phase accumulated after two interactions can be written as
    \begin{align}
        \frac{C\gamma}{\Delta+\delta}+\frac{C\gamma}{\Delta-\delta} = \frac{2C\gamma}{\Delta[1-(\delta/\Delta)^2]}
    \end{align}
    and the residual error resulted from the noise is then suppressed when $\delta/\Delta \ll 1$. This condition highlights that operating at a larger detuning $\Delta$ is also beneficial for mitigating dephasing noise.
    With the dynamical decoupling pulse, the relevant coherence limitation is no longer the $T_2^*$ but rather the $T_2$. For NV centers hosted in ultrapure diamond, $T_2$ exceeds $1~\mathrm{ms}$~\cite{Herbschleb2019}, which is far longer than the typical phase-encoding duration in our scheme. This is especially favorable because the optical lifetime of the excited state of NV centers is only $\tau_e \sim 12$~ns, enabling many rounds of photon-spin interactions. 
    Additionally, shorter duration reduces propagation loss in the optical fiber for NV-center implementations with the 637~nm photon. Finally, the coherence-time requirements emphasized here are not specific to NV centers or quantum dots. This applies broadly to solid-state spin-photon interfaces.
    
    \subsection{Optimal width-scaling scheme}
\begin{figure}[t]
        \centering
        \includegraphics[width=0.48\textwidth]{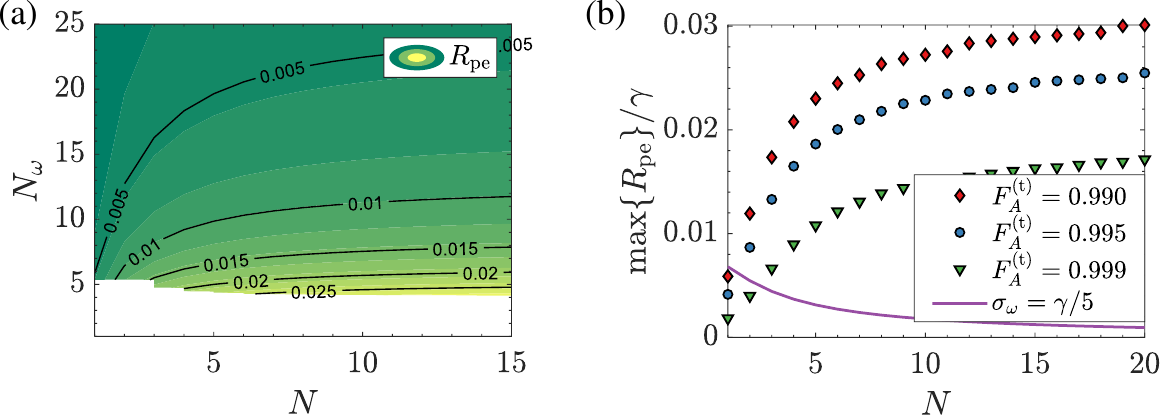}
        \caption{(a) The simulated success probability with different $N_\omega$ for each $N$. And the blank region has the fidelity achieved below the threshold $F_{A}^{(\mathrm{t})}=0.99$. (b) The maximal encoding rates under the fidelity thresholds $F_A^{(\mathrm{t})}=0.99,,0.995,$ and $0.999$ are shown by red diamonds, blue circles, and green triangles, respectively. Each data point corresponds to the optimal choice of $N_\omega$ for each $N$. The purple line shows the result for the width-fixed scheme with $\sigma_{\omega}=\gamma/5$.}
        \label{fig_SI_width_varying}
    \end{figure}

    In this section, we provide additional details on optimizing the width parameter $\Delta/N_\omega$. As a supplement to Fig.~\ref{fig_width_varying}, Fig.~\ref{fig_SI_width_varying}(a) shows the simulated encoding rates $R_{\mathrm{pe}}$ as a function of $N_\omega$ and the repetition number $N$. The blank region corresponds to the fidelity achieved below the threshold $F_{A}^{(\mathrm{t})}=0.99$. As we know, increasing $N_\omega$ narrows the pulse spectrum, which generally improves the port-$A$ fidelity and can also increase $P_t$ by better confining the pulse within the entangling window. Consequently, a given fidelity threshold $F_A^{(\mathrm{t})}$ can be reached by choosing sufficiently large $N_\omega$. However, a larger $N_\omega$ also lengthens the pulse in time and increases the phase-encoding duration $T_{\mathrm{pe}}$, so the encoding rate $R_{\mathrm{pe}}=P_t/T_{\mathrm{pe}}$ will be maximized at an optimal $N_\omega$ once the fidelity constraint is reached. By optimizing over $N_\omega$ for each fixed $N$, as shown in Fig.~\ref{fig_SI_width_varying}(b), we obtain the maximum achievable rates with the fidelity exceeds the thresholds $F_A^{(\mathrm{t})}=0.99,\,0.995,$ and $0.999$ are indicated by red diamonds, blue circles, and green triangles, respectively. As comparison, the result of the width-fixed scheme is presented by the purple line, which is much lower.

    \section{Effect of parameter variations}
    In the main text, we assume two identical transitions within each register and two identical registers. In practice, fabrication inevitably introduces parameter variations both within a register and between registers, which must be carefully accounted for.
    \label{Sec_SI_register_difference}
    
    \subsection{variations in one register}
    Here, we first discuss the effects of the different parameters in a single register $R_A$, where $g_{1,A}\neq g_{0,A}$ and $\gamma_{1,A}\neq\gamma_{0,A}$, leading to different cooperativities $C_{1,A}\neq C_{0,A}$ for the two transitions $\mathcal{T}_{1,A}$ and $\mathcal{T}_{0,A}$.
    For the monochromatic pulse, the spin-dependent reflection is
    \begin{align}
        r_{k,A} = 1- \frac{2\kappa_A \left(2i\Delta_{k,A} + \gamma_{k,A}\right)}{\kappa_A\left(2i\Delta_{k,A} + \gamma_{k,A}\right)+4g_{k,A}^2},
    \end{align}
    with the efficiency is 
    \begin{align}
        |r_{k,A}|^2  &= \frac{(C_{k,A}-1)^2+4\Delta_{k,A}^2/\gamma_{k,A}^2}{(C_{k,A}+1)^2+4\Delta_{k,A}^2/\gamma_{k,A}^2}\nonumber\\
        &=1-\frac{4C_{k,A}}{(C_{k,A}+1)^2+4\Delta_{k,A}^2/\gamma_{k,A}^2}.
        \label{eq_SI_r2}
    \end{align}
    As we can see, with large $C\rightarrow \infty$ and zero $C=0$ cooperativity, the reflection efficiency approaches 1, which enables the implementation of high-quality entanglement in the three-level $L$-configuration registers in the strong coupling regime~\cite{Nemoto2014}. And in the large detuning regime, with the increase of $\Delta_{k,A}/\gamma_{k,A}$, the reflection efficiency $|r_{k,A}|^2$ monotonically increases and gradually approaches 1.
    
    In single register, to achieve the entangling condition $(r_{0,A})^N+(r_{1,A})^N=0$, thus requiring two conditions: (i) $|r_{0,A}|=|r_{1,A}|$, and (ii) $N(\theta_{0,A}-\theta_{1,A})=(2m+1)\pi$ with $m=1$ is picked to minimize the repetition $N$.
    At first, condition (i) determines $\Delta_{1,A}$ as
    \begin{align}
        \Delta_{1,A} = -\frac{\gamma_{1,A}}{2} \sqrt{\frac{|r_{0,A}|^2(C_{1,A}+1)^2-(C_{1,A}-1)^2}{1-|r_{0,A}|^2}}.
    \end{align}
    This is valid if
    \begin{align}
        |r_{0,A}|^2 > \left(\frac{C_{1,A}-1}{C_{1,A}+1}\right)^2.
        \label{eq_SI_condition}
    \end{align}
    The reflection efficiency $|r_{0,A}|^2$~\eqref{eq_SI_r2} is monotonically increasing with detuning $\Delta_{0,A}$, which will make the condition~\eqref{eq_SI_condition} be always satisfied if the detuning $\Delta_{0,A}$ is larger than a value making two side of~\ref{eq_SI_condition} are equal. 
    And then the optimal detuning value $\Delta_{0,A}$ can be obtained in this regime to eventually accumulate a phase difference of $\pi$ with a corresponding different repetition $N$ required to fulfill condition (ii).
    
    Typically, the reflections with same amplitude $R_A(\omega)$ could be written as $r_{0,A}(\omega)=-R_A(\omega)\exp(i\theta_{0,A}(\omega))$ and $r_{1,A}(\omega)=-R_A(\omega)\exp(i\theta_{1,A}(\omega))$ with the phase $|\theta_{0,A}(\omega)|$ is no longer equal to $|\theta_{1,A}(\omega)|$. Then the final state of the spin and photon is 
    \begin{align}
        (-1)^N\int d\omega &\left[|0\rangle_A  e^{iN\theta_{0,A}(\omega)} + |1\rangle_A  e^{iN\theta_{0,A}(\omega)-\pi}\right]\times\nonumber\\
        &R_A(\omega)^N \tilde{u}(\omega) a^\dagger_A(\omega)|\mathrm{vac}\rangle.
        \label{eq_SI_phase_encoding_diff}
    \end{align}
    
    \subsection{variations in both registers}
    Generally, we assume that four transitions $\mathcal{T}_{s,M}$ of two registers $R_M$ having different parameters, which result in different reflections for each transition is
    \begin{align}
        r_{s,M}(\omega) = 1- \frac{2\kappa_M \left[\gamma_{s,M} + i2(\Delta_{s,M} - \omega)\right]}{(\kappa_M-2i\omega)[\gamma_{s,M}+2i(\Delta_{s,M}-\omega)]+4g_{s,M}^2}.
    \end{align} 
    
    Following the methods developed in Ref.~\cite{Omlor2025}, we could get the final joint state as
    \begin{widetext}
        \begin{align}
            |\psi_f\rangle = \frac{1}{\sqrt{2}}\int d\omega \Big\{&\left[\left(|\Phi^+\rangle+|\Psi^+\rangle\right) r_+^+(\omega)+|\Phi^-\rangle r_-^+(\omega) + |\Psi^-\rangle r_-^-(\omega)\right]\tilde{u}(\omega)a^\dagger_A(\omega)\nonumber\\
            +i&\left[\left(|\Phi^+\rangle+|\Psi^+\rangle\right) r_+^-(\omega)+|\Phi^-\rangle r_-^-(\omega) + |\Psi^-\rangle r_-^+(\omega)\right]\tilde{u}(\omega)a^\dagger_B(\omega)\Big\}|\mathrm{vac}\rangle.
        \end{align}
    \end{widetext}
    The parameters are defined as
    \begin{align}
        r^{\pm}_{\alpha}(\omega) = \frac{1}{2}[r_{\alpha,A}(\omega)\pm r_{\alpha,B}(\omega)]
    \end{align}
    with $\alpha=\{+,-\}$, and $r_{\pm,M}(\omega)=[r_{0,M}(\omega)^N\pm r_{1,M}(\omega)^N]/2$.
    Then the probabilities of getting a click on ports $A$ and $B$ are
    \begin{align}
        P_A &= \frac{1}{2} \int d\omega \left[2|r_+^+(\omega)|^2+ |r_-^+(\omega)|^2+|r_-^-(\omega)|^2\right]|\tilde{u}(\omega)|^2\nonumber\\
        P_B &= \frac{1}{2} \int d\omega \left[2|r_+^-(\omega)|^2+ |r_-^+(\omega)|^2+|r_-^-(\omega)|^2\right]|\tilde{u}(\omega)|^2\, .\nonumber\\
    \end{align} 
    And the fidelities with respect to two corresponding heralded Bell states $|\Phi^-\rangle$ and $|\Psi^-\rangle$ are
    \begin{align}
        F_{A/B} &= \frac{\int d\omega |r_-^+(\omega)|^2|\tilde{u}(\omega)|^2}{2P_{A/B}}.
    \end{align}
    In the identical case (i.e., four identical transitions), the final state is exactly the result~\eqref{eq_SI_pis_f_iden}. 
    Under the entangling condition for each register, $r_{+,M}(\omega) = 0$, making $r_+^{\pm}(\omega) =0$ and the resulting fidelities are 
    \begin{align}
        F_{A/B} &= \frac{\int d\omega |r_-^+(\omega)|^2 |\tilde{u}(\omega)|^2}{\int d\omega \left[|r_-^+(\omega)|^2+|r_-^-(\omega)|^2\right] |\tilde{u}(\omega)|^2}.
    \end{align}
    Here $r_-^-(\omega) = [r_{0,A}(\omega)-r_{0,B}(\omega)]/2$ represents the difference between two reflections of different registers. Thus, minimizing this directly enhances the fidelity. In addition, this also allows for a simple post-correction after the phase-encoding process if we could modify the reflection to make them be identical. 
    Once all parameters of the four transitions (i.e., $\gamma$, $g$, and $\kappa$) are known, we can insert an additional correction operation $U(\sqrt{r},\varphi/2)$ (see the blocks in Fig.~1(b) in the main text) in the optical path with the larger reflection amplitude, $\max(|r_{0,A}(0)|,|r_{0,B}(0)|)$, say path $A$ for example. Since this correction operation is placed in the optical path, the input photon in path $A$ passes through it once before and once after the multi-round phase-encoding process, so that the effective reflection amplitude of register $R_A$ could be transferred to close to that of $R_B$ in the entangling window as
    \begin{align}
        r_{0,B}(\omega)^N \approx U(\sqrt{r},\varphi/2)\,r_{0,A}(\omega)^N\,U(\sqrt{r},\varphi/2),
    \end{align}
    thereby maximizing the fidelities. The required parameters are given by the monochromatic limit 
    \begin{align}
        r = \left|\frac{r_{0,A}(0)}{r_{0,B}(0)}\right|^N, \quad 
        \varphi = N\left[\theta_{0,A}(0) - \theta_{0,B}(0)\right].
    \end{align}
    
\begin{figure}[t]
        \centering
        \includegraphics[width=0.49\textwidth]{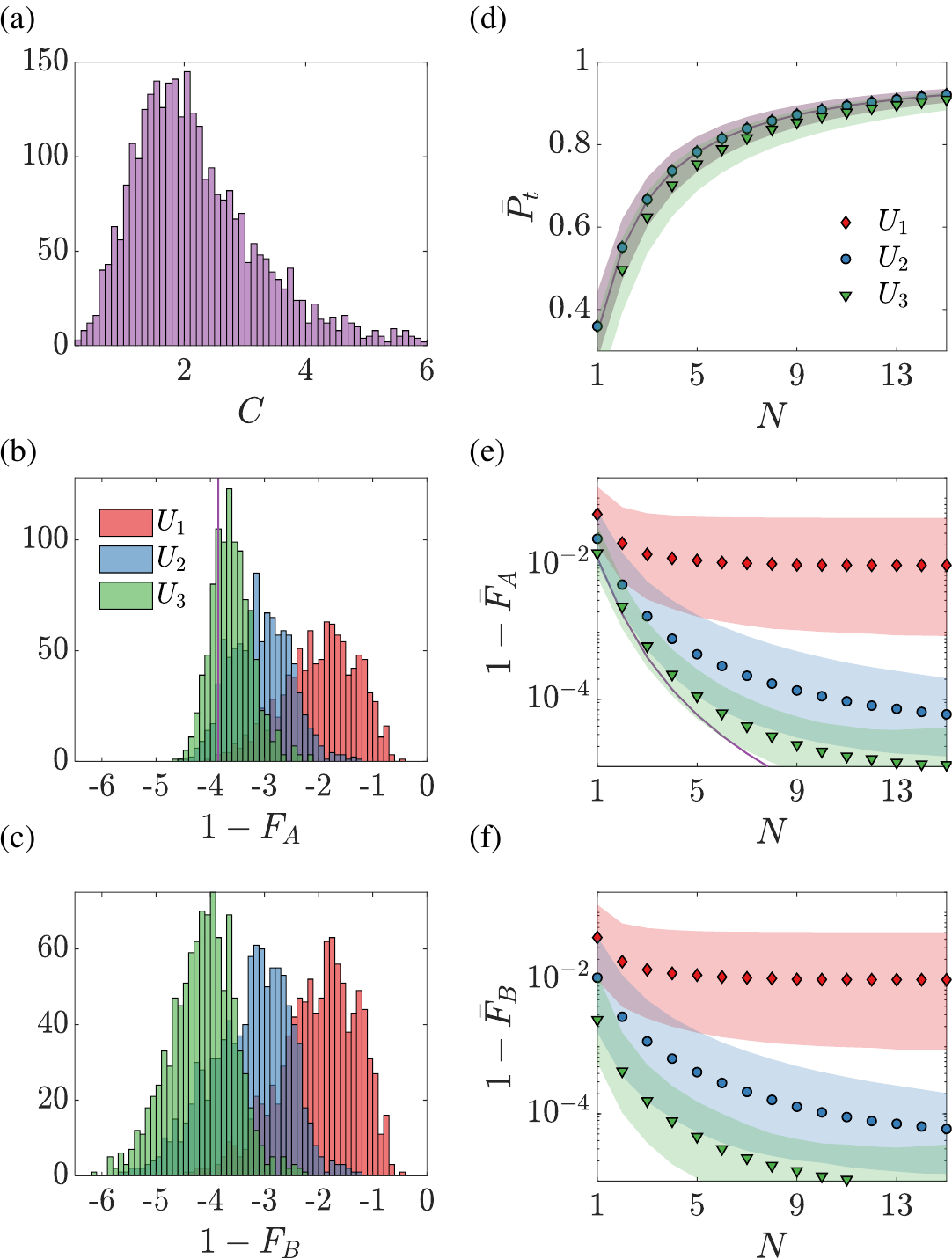}
        \caption{Simulation of four random transitions. (a) The histogram of the cooperativities sampled. (bc) The distributions of the simulated infidelities $1-F_A$ and $1-F_B$ with three different corrections used: $U_1$ red, $U_2$ blue, and $U_3$ green. Here, a negative integer $-m$ indicates a infidelity $10^{-m}$. (d--f) The averaged probabilities, infidelities of the entangling as a function of the number of interactions $N$ for the three different correction operations. The shaded bands indicate the region between the 16th and 84th percentiles.}
        \label{fig_SI_parameters}
    \end{figure}

    To further quantify the robustness of the entangling protocol against the parameter disorder, we perform a randomized disorder sampling of the cavity--spin parameters $\left(g,\gamma,\kappa\right)$ for all transitions $\{\mathcal{T}_{0,A},~\mathcal{T}_{1,A},~\mathcal{T}_{0,B},~\mathcal{T}_{1,B}\}$. We start from the reference transition $\mathcal{T}_{0,A}$ by setting the reference parameters $(g_{0,A},\gamma_{0,A},\kappa_{0,A})$ corresponding to a cooperativity $C_{0,A} = 2$ and a cavity loss $\kappa_0 = 200\gamma_0$. The remaining three transitions experience static relative fluctuations from reference values of $\mathcal{T}_{0,A}$. Specifically, for each disorder sample we draw the parameters sampled from $g_0 (1+\mathcal{N}(0,\sigma_g))$, $\gamma_0 (1+\mathcal{N}(0,\sigma_{\gamma}))$ and the loss for register $R_B$ is sampled from $\kappa_0 (1+\mathcal{N}(0,\sigma_{\kappa}))$.
    Here $\mathcal{N}$ is the normal distribution, and the relative amplitudes are set identically as $\sigma_{\gamma,\kappa,g} = \sigma$ for simplicity. This defines, for each sampling, a set of register parameters and the corresponding cooperativities. Their statistics, obtained from $M=1000$ independent disorder samples, are shown as in Fig.~\ref{fig_SI_parameters}(a) with $C_{0,A}=2$ and $\sigma=0.2$.
    
    In the numeric, we consider three different correction options: (i) no correction, $U_1=U(1,0)=\mathbbm{1}$; (ii) phase-only correction, $U_2=U(1,\varphi/2)$; and (iii) amplitude-and-phase correction, $U_3=U(\sqrt{r},\varphi/2)$. Then, for each value of $N$, we evaluate the resulting performance over the same set of disorder samples. Fig.~\ref{fig_SI_parameters}(b,c) show histograms of the infidelities $1-{F}_A$ and $1-{F}_B$ for the three correction operations $U_1$ (red bars), $U_2$ (blue bars), and $U_3$ (green bars). The integer $-m$ below the $x$ axis corresponds to an infidelity of $10^{-m}$. Furthermore, the purple line in Fig.~\ref{fig_SI_parameters}(b) indicates the reference infidelity $1-F_A$ for the ideal case of four identical transitions ($C=2$). This comparison shows that, with an appropriate correction, the entanglement fidelity is substantially enhanced and approaches the performance of the fully identical-transition case.
    
    Moreover, as illustrated in Fig.~\ref{fig_SI_parameters}(d), the average success probability $\bar{P}_t$ increases monotonically with $N$, while showing a weak dependence on the particular correlation of $U_1$, $U_2$, or $U_3$ and close to the reference probability of the l. In contrast, the fidelity is highly sensitive to the correction. Without any correction ($U_1$), the average infidelities $1-\bar{F}_A$ and $1-\bar{F}_B$ remain at the $\sim 10^{-2}$ level. Applying a phase-only correction ($U_2$) and amplitude-and-phase correction ($U_3$) will substantially suppress the infidelity and yield a clear improvement with $N$. The purple line in Fig.~\ref{fig_SI_parameters}(e) provides the reference infidelity $1-F_A$ for the ideal case of four identical transitions ($C=2$), and the $U_3$ results approach this benchmark, showing that an appropriate correction can largely restore the performance despite realistic parameter differences. The shaded bands with the same color indicate the region between the 16th and 84th percentiles. 
    
    \section{Phase Stability}
    \label{Sec_SI_phase}
    The successful heralding of a pure Bell state in our architecture depends critically on maintaining interferometric phase stability over two distinct paths A and B in the MZI. Compared to the traditional MZI-based single-shot entangling proposal, we employed an external loop to recycle the photon for $N$ times, which is expected to introduce additional phase errors. 
     
    Here we assume additional phase errors $\phi_A/N$ and $\phi_B/N$ arise on path $A$ and $B$ within each external loop, which will modify the reflection as $r_{s,M}\rightarrow r_{s,M}e^{i\phi_M/N}$. We could write the state~\eqref{eq_SI_state} after single  as 
    \begin{align}
        |\psi_1\rangle =& \frac{ia^\dagger_A} {2\sqrt{2}}e^{i\phi_A/N}\left(r_0|00\rangle+r_0|01\rangle+r_1|10\rangle+r_1|11\rangle\right)\nonumber\\
        +& \frac{a^\dagger_B} {2\sqrt{2}}e^{i\phi_B/N}\left(r_0|00\rangle+r_1|01\rangle+r_0|10\rangle+r_1|11\rangle\right).
    \end{align}
    This is simplified by assuming that $r_s = r_{s,A} = r_{s,B}$. As previously mentioned, this is permitted if we consider two identical registers or correct the reflection difference.
    
    Therefore, we get the final state as
    \begin{align}
        |\psi_f\rangle
        &= \frac{e^{i\bar\phi}}{\sqrt{2}}a_A^\dagger\left[\cos\left(\frac{\delta}{2}\right)|S\rangle+i\sin\left(\frac{\delta}{2}\right)r_-|\Psi^-\rangle\right]|\mathrm{vac}\rangle\nonumber\\
        &+ \frac{i e^{i\bar\phi}}{\sqrt{2}}a_B^\dagger\left[i\sin\left(\frac{\delta}{2}\right)|S\rangle+\cos\left(\frac{\delta}{2}\right)r_-|\Psi^-\rangle\right]|\mathrm{vac}\rangle,
    \end{align}
    with $\bar{\phi}=(\phi_A+\phi_B)/2$ and $\delta=\phi_A-\phi_B$.
    Here, the global phase $e^{i\bar\phi}$ is physically irrelevant while the relative phase $\delta$ mixes the state $|S\rangle$ and $|\Psi^-\rangle$ across the two output ports. In this case, we have the probabilities
    \begin{align}
        P_A &= \frac{|r_-|^2}{2}+|r_+|^2\cos^2\left(\frac{\delta}{2}\right)\nonumber\\
        P_B &= \frac{|r_-|^2}{2}+|r_+|^2\sin^2\left(\frac{\delta}{2}\right),
    \end{align}
    and the corresponding fidelities 
    \begin{align}
        F_A &= \frac{|r_-|^2\cos^2\left(\delta/2\right)}{|r_-|^2+2|r_+|^2\cos^2\left(\delta/2\right)}\nonumber\\
        F_B &= \frac{|r_-|^2\cos^2\left(\delta/2\right)}{|r_-|^2+2|r_+|^2\sin^2\left(\delta/2\right)}.
    \end{align}
   
    Note that the accumulated $N$-round cycle phase enters only as a multiplicative factor $e^{i\phi_M}$ on each arm and thus affects the protocol solely through the relative phase $\delta=\phi_A-\phi_B$ at the final BS. This is analogous to relative optical-path phase fluctuations between the two MZI arms, so the external-loop-induced and MZI-induced phase fluctuations can be absorbed into the same relative phase variable $\delta$. Hence, both contributions can be modeled by a single effective relative phase-noise parameter $\delta$.
        
    We assume the relative phase noise is Gaussian, $\delta=\delta_0+\delta(t)\sim\mathcal N(\delta_0,\sigma_\delta^2)$, with
    \begin{align}
        p(\delta)=\frac{1}{\sqrt{2\pi}\sigma_\delta}\exp\left[-\frac{(\delta-\delta_0)^2}{2\sigma_\delta^2}\right].
    \end{align}
    The constant (or nearly constant) phase offset, $\delta_0$, captures a static optical-path imbalance and very slow drifts, such as fixed fiber-length mismatch and thermal or strain variations that change on timescales of minutes to hours (i.e., <1 Hz). In contrast, the rest fast term, denoting by $\delta(t)$, describes residual stochastic phase fluctuations within or between entanglement attempts. These fluctuations are typically caused by mechanical vibrations of the setup, laser phase noise, and so on. Depending on the implementation and environment, the relevant spectral content ranges from Hz to 100 kHz~\cite{Stolk2024, Stolk2025}.
    
    For simplicity, we study the impact of phase difference error on the entangling proposal by using the average fidelity under the entangling condition $r_+=0$. Now the average of the quality is
    \begin{align}
        \bar{X}=\langle X(\delta)\rangle=\int d\delta X(\delta)p(\delta).
    \end{align}
    The probabilities are unaffected ${P}_{A/B}=|r_-|^2/2$, which is very similar with the dephasing on a two-level system, where the phase fluctuation will not change the populations. 
    With respect to the fidelities, we have
    \begin{align}
        \bar{F}_A = \bar{F}_B = \frac{1}{2}[1+\langle\cos(\delta)\rangle] = \frac{1}{2}\left[1+\cos(\delta_0)e^{-\sigma_\delta^2/2}\right],
    \end{align}
    where the constant basis $\delta_0$ could be very large, for example, a 1 mm length difference ($\sim\lambda$) will totally shift the phase, and a larger $\delta_\sigma$ will ruin the state to a maximal mixed state. 
    
    In an active phase-stabilized scheme, $\delta_0$ acts as a deterministic bias that is calibrated into a chosen setpoint $\delta_{\mathrm{set}}$, leaving a residual offset $\tilde{\delta}_0=\delta_0-\delta_{\mathrm{set}}$. For short fiber links, $\delta_{\mathrm{set}}$ can remain effectively constant over several attempts and thus need not be updated every time~\cite{Stolk2024}. Operationally, $\delta_{\mathrm{set}}$ is obtained by co-propagating a stabilized classical reference field with the quantum photon through the same optical paths, estimating the differential phase, and applying feedback to impose a compensating phase shift so that the relative phase is locked to the desired setpoint~\cite{Stolk2024}. This deterministic correction can be absorbed into a local phase rotation on one spin or, equivalently, into the phase shifter already used to compensate fixed register-to-register reflection differences. Consequently, the slow offset is suppressed to $\tilde{\delta}_0\sim0$, while stochastic fluctuations are reduced within the correction bandwidth, leaving $\sigma_\delta\ll 1$ and thus
    \begin{align}
        \bar{F}_{A/B}\approx1-\frac{\delta_{\mathrm{rms}}^2}{4}.
    \end{align}
    with $\delta_{\mathrm{rms}}=\sqrt{\tilde{\delta}_0^2+\sigma_\delta^2}$ is the root mean square (rms) of the residual relative phase error.
    
    Moreover, we treat the phase as approximately constant across a single photon wavepacket, since fiber-induced path-length noise is typically slow (Hz--kHz) and thus varies negligibly over the pulse duration of tens of nanoseconds. Accordingly, the relative phase error in each attempt is modeled as a constant parameter, with $\delta(t)$ understood as the accumulated relative phase over the attempt duration.
    
    Experimentally, at the few-kilometer scale, state-of-the-art phase-stabilized fiber links can already reach very small residual phase noise. For example, Ref.~\cite{Nardelli2025} reports a residual relative phase error $\delta_{\mathrm{rms}}\approx 0.06$--$0.085$~rad ($3.5^\circ$--$4.9^\circ$), implying a fidelity above $0.998$ with a $2.1$~km phase-stabilized link.
    Here, we model the incorporation of our repeated phase-encoding as an additional process within the same interference apparatus to assess its impact on fidelity. The key point is that the external cycle effectively increases the MZI arm length to $L_{\mathrm{MZI}}+N\ell_{\mathrm{cycle}}$. This extension is compatible with standard co-propagating phase referencing, since a detuned classical stabilization field can traverse the same fiber path as the quantum photon while remaining far off-resonant from the cavity and thus not entering it. In the width-fixed scheme $\sigma=\gamma/5$ for four-level QDs, we have $\ell_{\mathrm{cycle}}=10$~m, so $N=10$ adds only $N\ell_{\mathrm{cycle}}\approx100$~m, much smaller than a few-kilometer MZI arm.
    When the phase stabilization operates in the delay-limited regime for long fiber links, and the fiber phase noise can be modeled as spatially distributed and uncorrelated along the fiber link, the free-running one-way phase-noise spectrum scales as $S_{\mathrm{fiber}}(f)\propto L/f^2$, where $L$ is the fiber length. Under active phase stabilization, the residual phase error obeys $S_{\mathrm{lock}}(f)\propto (2\pi f\tau)^2 S_{\mathrm{fiber}}(f)$, with $\tau=L/c_n$ the one-way propagation delay and $c_n$ the speed of light in the fiber. This scaling implies $\delta_{\mathrm{rms}}\propto L^{3/2}$ for the residual one-way phase fluctuations~\cite{Williams2008, Nardelli2025}. For two equal-length and long fiber links, statistically independent links $A$ and $B$, the relative phase $\phi_A-\phi_B$ exhibits the same scaling $\delta_{\mathrm{rms}}\propto L^{3/2}$. This relation provides a simple order-of-magnitude estimate of the length dependence of $\delta_{\mathrm{rms}}$.
    
    As a result, adding $N\ell_{\rm cycle}=100$~m changes $2.1$~km to $2.2$~km and yields a predicted fidelity $0.9979$, still well above $0.99$. At the 10-km scale for remote entanglement, the residual relative phase error can be suppressed to $\delta_{\mathrm{rms}}\approx 0.61,\,0.30$ ($35^\circ,\,17^\circ$)~\cite{Stolk2024, LiuJL2024}, thereby supporting the feasibility of our proposal in a metropolitan quantum network. Moreover, an even smaller $\ell_{\mathrm{cycle}}$ is used in the width-scaling scheme, enabling $N\ell_{\mathrm{cycle}}\ll L_{\mathrm{MZI}}$ and thus only a negligible reduction of the fidelity.
    
    In the other side, considering two close registers, i.e., $N\ell_{\mathrm{cycle}}\gg L_{\mathrm{MZI}}$, a $N=200$ proposal with $F_A\sim0.998$ is ensured in this 2-km setup. In addition, co-routing the two fibers through the same path and environment can substantially suppress the residual relative phase error via the common-mode noise effect, further improving the fidelity. Overall, these considerations indicate that our scheme remains fully compatible with conventional one-shot entanglement architectures and can achieve high-fidelity operation using established, standard phase-stabilization techniques.
    
    \section{The loss channels}
    \label{Sec_SI_loss}
    In this section, we analyze the dominant photon-loss mechanisms in the optical setup, namely the intrinsic loss of the cavity, the insertion loss associated with optical components, and the propagation loss along the optical path. In our scheme, the overall photon reflection directly determines the achievable entanglement generation: smaller optical loss leads to a higher success probability. In contrast, provided that the protocol remains within the intended operational regime (in particular, that the entangling condition is satisfied), these loss channels predominantly reduce the heralding rate rather than introducing a significant additional change in the heralded states. Therefore, we focus here on quantifying their impact on the success probability, which is mainly relevant to the last part of the section of protocol optimization in the main text.
    
    For the sake of simplicity, we assume four identical transitions and operate under the entangling condition, such that the spin-dependent cavity reflection realizes the desired spin-photon phase encoding. In this entangling window $r_0(\omega)\approx r_1(\omega)$, the total success probability can be approximated as
    \begin{align}
        P_t = \int d\omega |r_0(\omega)|^2 |u(\omega)|^2.
    \end{align}
    Accordingly, the problem reduces to characterizing how each loss contribution changes the reflection coefficient $r_0(\omega)$, from which the corresponding reduction of $P_t$ follows immediately. 
    In our analysis, we will deal with the monochromatic input pulse, yielding the total entangling probability is equal to the reflection probability $P_t\approx|r_0(\omega=0)|^{2N}$. While in the simulation through this section, we will use the Gaussian pulse with a fixed width $\sigma_{\omega}=\gamma/5$ and $\kappa/\gamma=200$.
    
    \subsection{The cavity loss}
    \label{Sec_SI_cavity_loss}
    One of the dominant loss channels in our optical setup is the cavity loss. In an ideal scenario of a cavity, the photon is trapped in the cavity and undergoes repeated reflections from the mirrors, thereby enhancing the effective spin-photon interaction. In practice, however, unwanted photon loss must be carefully accounted for. We model the total cavity energy-decay rate as
    \begin{align}
        \kappa=\kappa_l+\kappa_r+\kappa_i
    \end{align}
    where $\kappa_l$ and $\kappa_r$ denote the decay rates through the left and right cavity mirrors, and $\kappa_i$ captures the intrinsic loss in the cavity. In our proposal, we assume we have a single-sided cavity with $\kappa_r=0$ and also an overcoupled cavity $\kappa\gg\kappa_i$ to ensure the photon will be guided to the left side of the cavity. Here we define a cavity efficiency as $\eta_i=\kappa_l/(\kappa_l+\kappa_i)$, and as we assumed in the main text, $\eta_i=1$, resulting in a perfect reflection. Considering the cavity loss, the reflection coefficient is
    \begin{align}
        r_0(\eta_i) = 1- \frac{2\eta_i\kappa \left(\gamma +2i \Delta_0\right)}{\kappa\left(\gamma + 2i\Delta_0\right)+4g^2}.
    \end{align}
    With $\eta_i\rightarrow1$, we have the ratio
    \begin{align}
        \frac{r_0(\eta_i)}{r_0(1)}=1+ \frac{2\kappa \left(\gamma + 2i\Delta_0\right)}{4g^2-\kappa\left(\gamma + 2i\Delta_0\right)}(1-\eta_i).
    \end{align}
    In the large detuning limit $\Delta_0/\gamma\rightarrow \infty$, ignoring the tiny change of the phase, we could write the modified reflection 
    \begin{align}
        r_0(\eta_i)\approx  \eta_i^2r_0(1),
    \end{align}
    since $1-2\eta_i+\eta_i^2 = (1-\eta_i)^2\sim 0$ in the larger detuning case.
    \begin{figure}[t]
        \centering
        \includegraphics[width=0.48\textwidth]{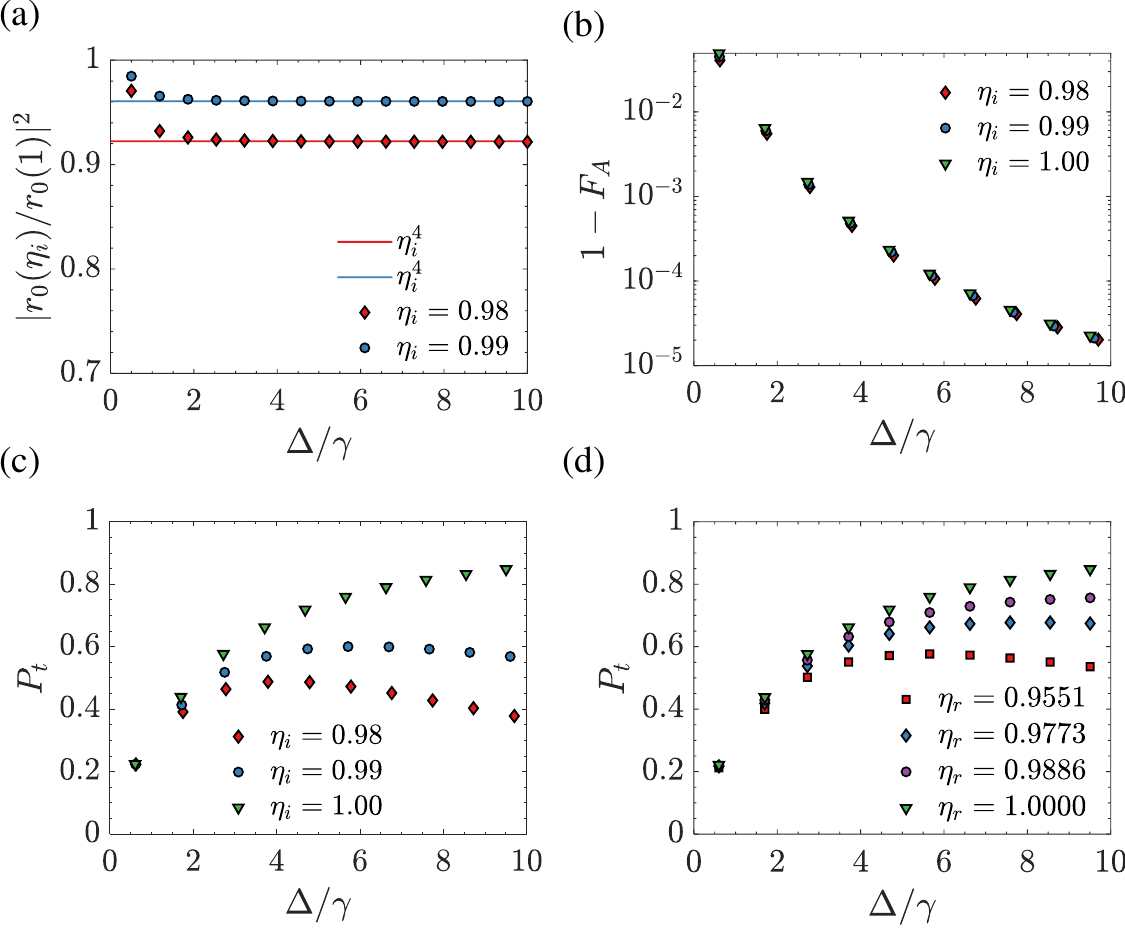}
        \caption{(a) Ratio of the reflection probability for cavity efficiencies $\eta_i=0.99$ and $0.98$ (blue circles and red diamonds) relative to the ideal case $\eta_i=1$, i.e., $|r_0(\eta_i)|^2/|r_0(\eta_i=1)|^2$. The lines with the same color show the values of $\eta_i^4$. (b,c) Simulated entangling fidelities and success probabilities for different cavity efficiencies, evaluated at the optimal detuning values $\Delta$, for $\eta_i=0.98, 0.99,$ and $1$ (red diamonds, blue circles, and green triangles). The fidelities vary only weakly, whereas the success probability decreases as $\eta_i$ decreases. The success probability is maximized at an optimal detuning $\Delta$, and the optimal detuning point shifts slightly because $\eta_i$ modifies the reflection coefficient $r_0$. of 0.2 dB($\eta_r=0.9551$, red squares), 0.1 dB($\eta_r=0.9773$, red squares), 0.05 dB($\eta_r=0.9886$, red squares), and 0 dB($\eta_r=1$, red squares). Parameters $C=1.5$.}
        \label{fig_SI_loss}
    \end{figure}
    
    In Fig.~\ref{fig_SI_loss}(a), we plot the ratio $|r_0(\eta_i)|^2/|r_0(\eta_i=1)|^2$ for $\eta_i=0.98$ and $0.99$ (red diamonds and blue circles), showing good agreement with the values of $\eta_i^4$. 
    Moreover, we simulate the entangling protocol in the presence of cavity loss for cooperativity $C=1.5$. Fig.~\ref{fig_SI_loss}(b,~c) present the simulated entangling fidelities and success probabilities at the optimal detuning values $\Delta$ for $\eta_r=0.98$, $0.99$, and $1$ (red diamonds, blue circles, and green triangles). The fidelity $F_A$ depends only weakly on the loss, since the cavity loss only has a tiny effect on the phase of the reflection, leading to only a slight shift of the optimal detuning point. By contrast, the success probability decreases as $\eta_i$ decreases since more and more photons leak out of the cavity through the loss channel $\kappa_i$. By contrast, the success probability decreases as $\eta_i$ decreases, since an increasing probability of photons leaks out through the intrinsic-loss channel $\kappa_i$. For each $\eta_i$, the success probability is maximized at one specific optimal detuning.
    
    It should be noted that in actual experiments, the intrinsic loss of the cavity is highly dependent on the manufacturing quality of the cavity. The intrinsic loss provides a link between the per-round-trip loss and the cavity finesse. The intrinsic loss originates predominantly from absorption and scattering processes. For fiber-based or conventional FP microcavities, the leading contributions to $\kappa_i$ arise from (i) absorption in the dielectric mirror coatings, (ii) scattering from surface roughness and coating inhomogeneities on the mirror surfaces, and (iii) additional absorption or scattering from intracavity media. For an NV center coupled to an open cavity~\cite{Yurgens2024}, the intrinsic loss per round trip has been measured to be on the order of several tens to several hundreds of ppm per round trip. This loss is expected to be significantly reduced with improved diamond surface quality (reduced roughness) and higher-quality cavity mirrors. In addition, subwavelength nanodiamond-hosted NV centers may reduce undesired scattering into non-cavity channels. Moreover, a larger value of $\kappa_l$ can also effectively reduce the impact of cavity loss. As an example, in a microcavity designed to couple with a quantum dot system, the transmission loss per round trip of the one mirror is 10300 ppm, which is much larger than intrinsic losses of about 373 ppm per round-trip, resulting in an efficiency $\eta_i\approx0.97$~\cite{Tomm2021}.
    
    \subsection{The optical loss}
    \label{Sec_SI_optical_loss}
    On the other hand, we have to consider the propagation loss along the optical path as well as the insertion loss of optical components such as optical switches. The power efficiency is denoted as $\eta_r$, with the effective now written as
    \begin{align}
        r_0^{\mathrm{eff}} \approx {\eta_r}^{1/2}\eta_i^2r_0.
    \end{align}    
    Fig.~\ref{fig_SI_loss}(d) shows the success probability for optical losses of 0.2~dB ($\eta_r=0.9551$), 0.1~dB ($\eta_r=0.9773$), 0.05~dB ($\eta_r=0.9886$), and 0~dB ($\eta_r=1$) for $C=1.5$.  
    To greatly increase the efficiency $\eta_r$, we suggest that the whole system is working at the telecom frequency. Thus, we could utilize the single-mode, low-loss fiber as the external route for the additional cycle and also employ the low-loss optical switch developed and optimized for the telecom band. 
    
    As an example, in semiconductor quantum-dot systems, the emission frequency has been tuned into the telecom band~\cite{Vajner2024, Yu2023}, and remote entanglement has been demonstrated using an emission-based protocol~\cite{Laccotripes2024}. For a quantum dot in the Voigt geometry, the excited-state lifetime is about $1$ ns, corresponding to a radiative decay rate $\gamma \approx (2\pi)160$ MHz. Considering a Gaussian input pulse with spectral width $\sigma_\omega=\gamma/5$, the temporal width is $\sigma_t= 1/\sigma_\omega$, so a pulse length $T_p=10\sigma_t=10/\sigma_\omega\approx 50$~ns is sufficient(a shorter pulse duration is discussed in \ref{Sec_SI_duration}). This corresponds to an additional delay-line length of only $\approx 10$ m of fiber per cycle. For low-loss telecom fiber with an attenuation of $\sim 0.2$~dB/km, the propagation loss over $10$~m can be safely ignored for moderate values of $N$.
    
    In addition, the PsiQuantum team reported a low insertion loss of about $0.1$ dB for a $2\times 2$ optical switch operating at GHz rates~\cite{PsiQuantum2025}. This would enable a fast enough $3\times 1$ optical switch by cascading two such $2\times 2$ devices. The resulting total insertion loss is then $0.2$ dB (i.e., $\eta_r=0.9551$). Their further goal is to halve this loss to $0.1$ dB (i.e., $\eta_r=0.9773$). Optical insertion loss does not shift the optimal detuning values, but it reduces the success probability.
    
    \begin{figure}[t]
        \centering
        \includegraphics[width=0.48\textwidth]{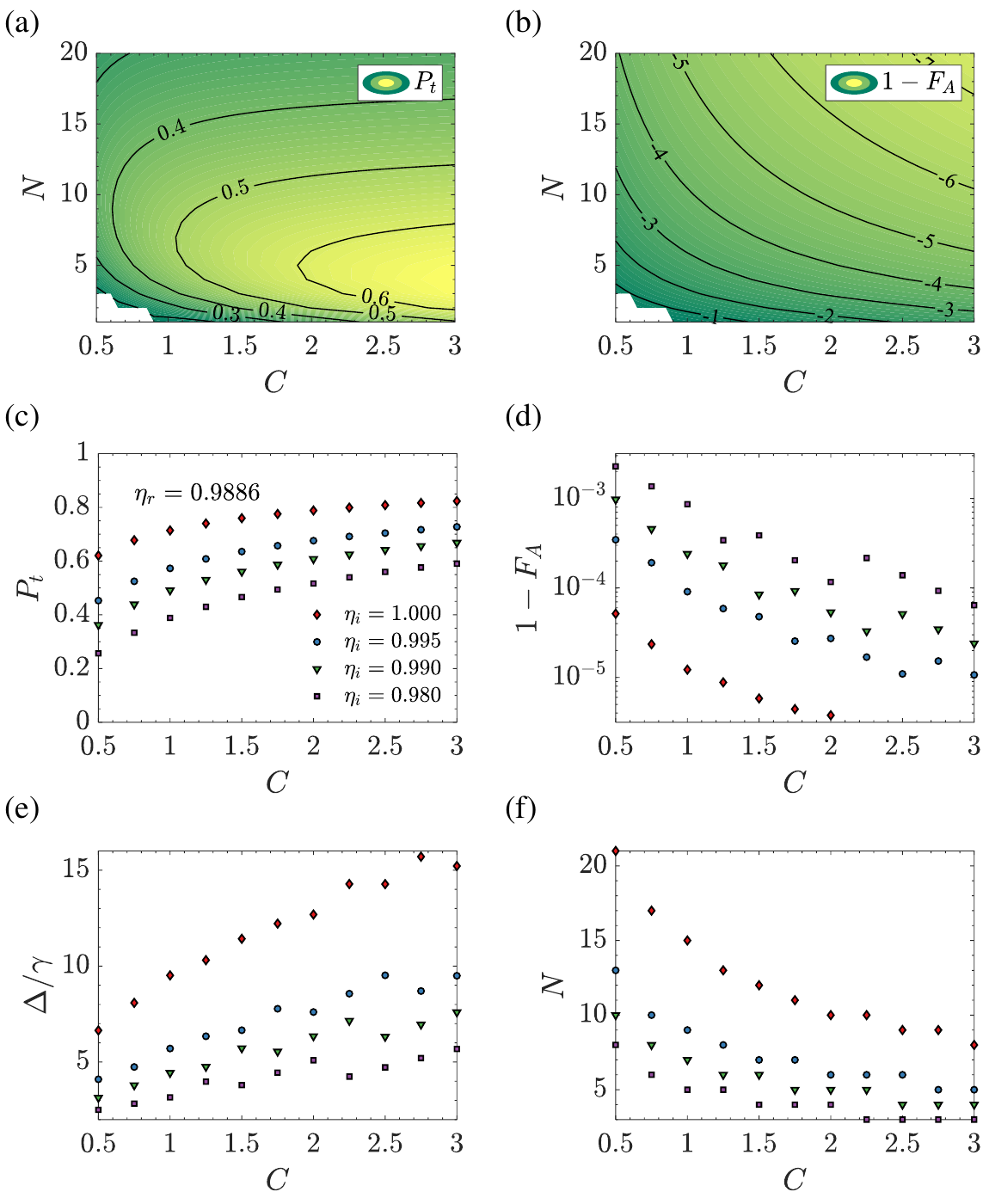}
        \caption{Simulation of the loss effect. (a,b) The simulated probability $P_t$ and fidelity $F_A$ as a function of cooperativity $C$ and repetition $N$ with the parameters are $\eta_i=0.99$ and $\eta_r=0.9886$. (c-f) show the simulated maximal probability $P_t$, fidelity $F_A$, optimal detuning values $\Delta/\gamma$ and repetition $N$ for different cooperativity with $\eta_r=0.9773$. Here, the red diamonds, blue circles, green triangles and purple squares represent the results for $\eta_i=1,0.995,0.99,0.98$, respectively.}
        \label{fig_SI_cavity_round_loss_II}
    \end{figure}
    
    With increasing repetition number $N$, the total success probability scales as $P_t \approx |r_0^{\mathrm{eff}}|^{2N}$ and can be approximated by $(\eta_r \eta_i^4)^N |r_0|^{2N}$. Therefore, increasing $N$ entails two competing effects on probability. On the one hand, $|r_0|^2$ can be enhanced by suppressing spontaneous emission (i.e., reducing the effective impact of $\gamma$) using a large detuning $\Delta=2NC\gamma/\pi$. On the other hand, photon loss accumulates over repeated interactions and reduces the overall success probability. Consequently, one should choose an optimal repetition number $N$ that maximizes the success probability while maintaining an acceptable fidelity. In Fig.~\ref{fig_SI_cavity_round_loss_II}(a,b), we show the simulated probability and fidelity of the entangling for different cooperativity $C$ and repetition $N$ with $\eta_r=0.9986$ and $\eta_i=0.99$. As can be seen from the figure, in the presence of both cavity loss and optical loss, the success probability is maximized at an optimal repetition number of approximately $N\approx 5$. 
    
    Fig.~\ref{fig_SI_cavity_round_loss_II}(c,d) show the maximal $P_t$ and the corresponding fidelity obtained at the optimal $N$ as a function of cooperativity. Here we fix $\eta_r=0.9773$ (0.1~dB loss), and the results for $\eta_i=1$, $0.995$, $0.99$, and $0.98$ are indicated by red diamonds, blue circles, green triangles, and purple squares, respectively. The corresponding optimal detuning $\Delta$ and repetition number $N$ are shown in Fig.~\ref{fig_SI_cavity_round_loss_II}(e,f) using the same markers. Increasing the cooperativity $C$ and improving the efficiencies $\eta_i$ and $\eta_r$ enhance both the success probability and the fidelity. This improvement arises because larger $C$ and higher $\eta_{i,r}$ reduce the optimal repetition number $N$ (Fig.~\ref{fig_SI_cavity_round_loss_II}(e)), which mitigates accumulated photon loss and thus increases $P_t$, and simultaneously shift the optimal detuning point to larger detuning $\Delta$ (Fig.~\ref{fig_SI_cavity_round_loss_II}(f)), which broadens the entangling window and enables higher fidelity.
    
    By improving fabrication quality and thereby further reducing losses in both the cavity and the optical path, one can realize phase encoding via additional rounds of spin-photon interactions and increase the entanglement success probability, which scales as $\sim 1-P_{\mathrm{loss}}$ with the photon-loss probability. Although the longer protocol duration reduces the entanglement generation rate, this strategy provides a practical route toward a photon-loss-tolerant architecture for quantum computing with moderate cooperativity; in typical spin-photon architectures, the total optical loss should be kept below the $\sim 0.1$ level. Moreover, our approach is naturally compatible with spin-photon hybrid quantum information processing, including fusion-based quantum computing with registers containing multiple spin qubits. In such platforms, the preparation of the required intra-register entangled resource states can take substantially longer than a single inter-register entangling attempt. Increasing the per-attempt success probability, therefore, reduces the number of repetitions needed to establish the desired links, lowering the time-to-solution and the photon budget.
    
    \section{Mode mismatch}
    \label{Sec_mode_mismatch}
    In this section, we are going to discuss the effect of mode match between the input optical mode and the eigenmode (usually the TM$_{00}$ mode). In an FFPC interface, imperfect mode matching between the fiber mode and the cavity mode is unavoidable. It can arise from residual lateral and angular misalignment, wavefront mismatch due to imperfect mode matching optics, polarization-dependent coupling, and slow drifts of the fiber--cavity geometry. A standard and experimentally measurable parameter is the mode-overlap efficiency $\eta_m$ between the outgoing field and the single-mode fiber mode.
    
    Here, we model the spatial mode of a single photon in the single-mode fiber by the normalized state $|f\rangle=a_f^\dagger|\mathrm{vac}\rangle$. Due to imperfect overlap, the input can be decomposed into a cavity-matched component $|m\rangle=a_m^\dagger|\mathrm{vac}\rangle=\int d\omega\tilde{u}(\omega) a^\dagger_{m}(\omega)|\mathrm{vac}\rangle$ and an orthogonal unmatched component $|n\rangle=a_n^\dagger|\mathrm{vac}\rangle$ as
    \begin{align}
        |f\rangle = \sqrt{\eta_m}|m\rangle + \sqrt{1-\eta_m}|n\rangle\, .
    \end{align}
    Only the matched component enters the cavity and acquires the spin-dependent reflection coefficient $r_M(\omega)$. Here, we have to emphasize that the mode $a_{f,m,n}^\dagger$ indicates the spatial mode of the fiber and cavity. The unmatched mode does not couple to the cavity and is promptly reflected with a spin-independent amplitude and phase $r_n e^{i\theta_n}$.  For a monochromatic pulse, the joint state after reflection, but before the fiber reads
    \begin{align}
        |\Psi\rangle &= \frac{1}{\sqrt{2}}\left(r_0|0\rangle|m\rangle + r_1|1\rangle|m\rangle\right)\sqrt{\eta_m}\nonumber\\
        &+\frac{1}{\sqrt{2}}\left(|0\rangle+|1\rangle\right)|n\rangle\sqrt{1-\eta_m}r_n e^{i\theta_n}
        \label{eq_SI_psi_ref}
    \end{align}
    with $r_M=r_M(\omega=0)$.
    
    The single-mode fiber after the reflection plays the role of a spatial mode filter, since only the mode $a_f^\dagger$ will be propagated by the fiber, while the unmatched mode will be blocked. Operationally, the fiber projects the outgoing field onto the fiber mode with the projector $\Pi_f = |f\rangle\langle f|$. Using $\langle f|m\rangle=\sqrt{\eta_m}$ and $\langle f|n\rangle=\sqrt{1-\eta_m}$, the post-filter state in the fiber mode is
    \begin{align}
        (\openone\otimes\Pi_f)|\Psi\rangle
        =\frac{1}{\sqrt{2}}\left(\tilde r_0|0\rangle+\tilde r_1|1\rangle\right)|f\rangle,
    \end{align}
    where the mode mismatch modifies the reflection coefficients to the effective amplitudes
    \begin{align}
        \tilde r_s = \eta_m r_s + (1-\eta_m) r_n e^{i\theta_n}.
        \label{eq_SI_r_im}
    \end{align}
    Similarly, after the $N$-th round, the state is
    \begin{align}
        \frac{1}{\sqrt{2}}\left[(\tilde r_0 )^N|0\rangle+(\tilde r_1)^N|1\rangle\right]|f\rangle.
        \label{eq_SI_f}
    \end{align}
    The effective reflection $\tilde r_s$ then changes in both magnitude and phase, thereby moving the optimal detuning and repetition number required by the phase-encoding condition, $(\tilde r_0 )^N+(\tilde r_1)^N=0$. Then the probability of success in this case is given by
    \begin{align}
        P_t\approx|\tilde{r}_0|^{2N} &= \big[R^2\eta_m^2+(1-\eta_m)^2r_n^2\nonumber\\&-2R\eta_m(1-\eta_m)r_n\cos\theta_n\cos\theta\big]^N,
    \end{align}
    where the last term involving $\eta_m(1-\eta_m)$ is the interference between the reflected matched and unmatched modes. Here we use $r_0=-Re^{i\theta}$ and $r_1=-Re^{-i\theta}$.

    We now discuss three experimentally relevant strategies for dealing with the unmatched component and their consequences for the effective reflection and photon loss.
    
    (i) No compensation. If one does nothing, the phase picked by the non-entering mode is $\theta_n=0$ according to the input-output relation(see Sec.~\ref{Sec_SI_input_out}) and $r_n=1$, giving an effective corresponding reflection 
    \begin{align}
        \tilde{r}_s^{\mathrm{id}} = \eta_m r_s + (1-\eta_m).
    \end{align}
    The probability of success is
    \begin{align}
        P_t^{\mathrm{id}}\approx\left[R^2\eta_m^2  + (1-\eta_m)^2 - 2R\eta_m(1-\eta_m)\cos\theta\right]^N.
    \end{align}
    Here, the last term is negative and refers to a destructive interference between matched and unmatched modes to enhance the photon loss.
    
    (ii) Mode separation. If the promptly reflected unmatched component is spatially separated and prevented from re-coupling back to the single-mode fiber before the fiber projection, then the fiber only receives the matched component (i.e., $r_n=0$). In this idealized case, the field incident on the fiber is proportional to $\sqrt{\eta_m}r_s|m\rangle$ and yields
    \begin{align}
        \tilde r_s^{\mathrm{sep}}=\eta_m r_s\, .
    \end{align}
    This makes the discarding of unmatched modes is treated as pure loss, like the optical loss. giving a probability
    \begin{align}
        P_t^{\mathrm{sep}}\approx |R^2\eta_m^2|^N.
    \end{align}
    Moreover, this is naturally fulfilled in the entangling protocol based on the transmitted photon, as we discussed in \ref{Sec_SI_transmission}.
    
    (iii) Selective $\pi$-phase compensation on the unmatched or matched mode. Instead of discarding $|n\rangle$, one can engineer its relative phase so that it interferes positively upon fiber filtering $r_n=1$ and $\theta_n=\pi$, yielding
    \begin{align}
        \tilde{r}_s^{\mathrm{sel}} = \eta_m r_s - (1-\eta_m),
    \end{align}
    with
    \begin{align}
        P_t^{\mathrm{sel}}\approx\left[R^2\eta_m^2  + (1-\eta_m)^2 + 2R\eta_m(1-\eta_m)\cos\theta\right]^N.
    \end{align}
    In the large-detuning regime where $R\simeq1$ and $\theta$ is small, this choice makes the total success probability approaching unit $P_t^{\mathrm{sel}}\rightarrow1$, this is because in this case, the reflected state \eqref{eq_SI_psi_ref} becomes close to the original fiber mode $|f\rangle$ and the additional loss induced by fiber filtering is strongly suppressed. 
    
    \begin{figure}[t]
        \centering
        \includegraphics[width=0.48\textwidth]{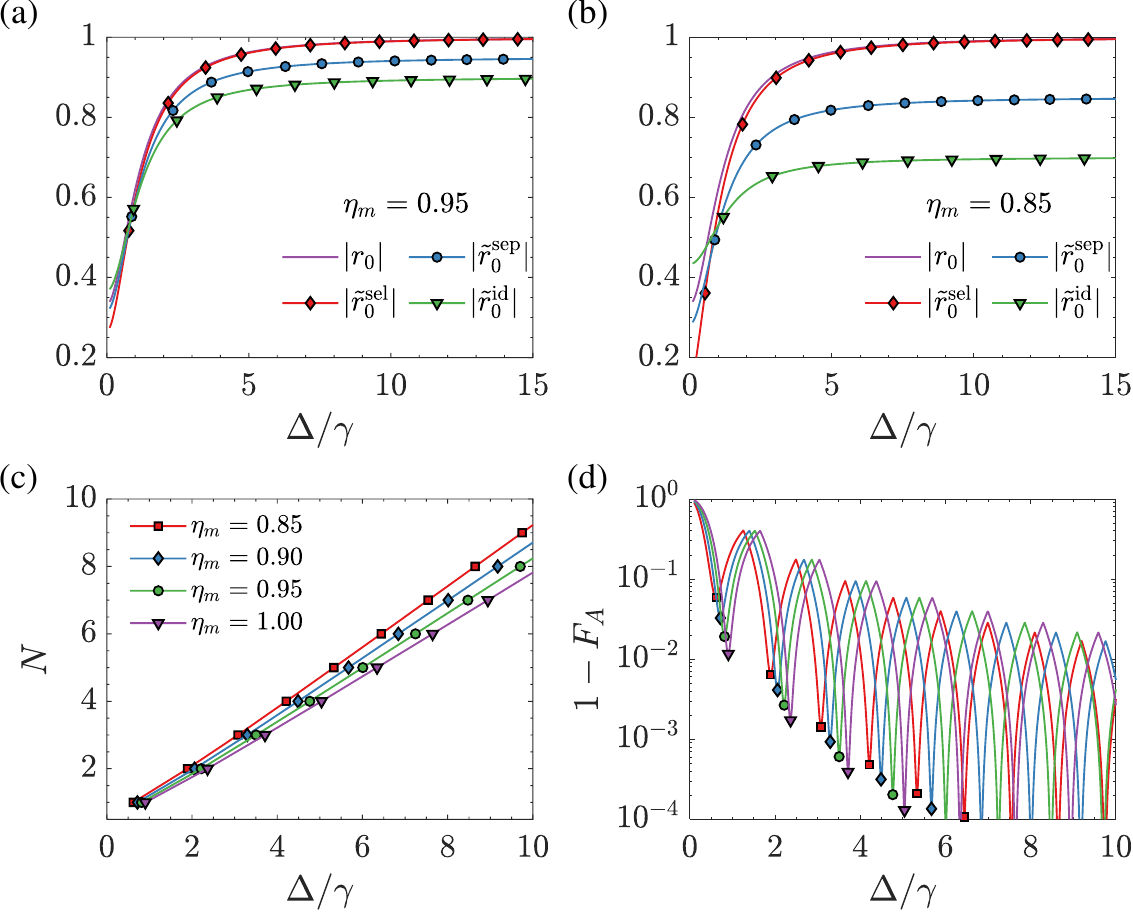}
        \caption{The reflection amplitudes of $|\tilde{r}_0^{\mathrm{sel}}|$, $|\tilde{r}_0^{\mathrm{sep}}|$, and $|\tilde{r}_0^{\mathrm{id}}|$ are denoted by red diamonds, blue circles and green triangles, respectively. The discrete symbols are the optimal values of detuning $\Delta$ to make $\arg(-\tilde{r}_0)=\pi/2N$. The corrected $|\tilde{r}_0^{\mathrm{sel}}|$ approaches the reference purple line($\eta_m=1$), demonstrating the effectiveness of the correction, and releasing the requirement of the high mode overlap $\eta_m$. Here, we choose $C=2$.}
        \label{fig_SI_mismatch_I}
    \end{figure}
    
    This could also be verified by minimizing the photon loss induced by fiber filtering. Here, we consider the first round of interaction, before the fiber filtering, the total reflected photon number (the photon number of state \eqref{eq_SI_psi_ref}) is
    \begin{align}
        P_{\mathrm{pre}} =& \frac{\eta_m}{2}\left(|r_0|^2+|r_1|^2\right) + (1-\eta_m)|r_n|^2 \nonumber\\
        =& R^2\eta_m+1-\eta_m\, .
    \end{align}
    After the fiber filtering, the photon number that remains in the single-mode fiber (state \eqref{eq_SI_psi_f}) is
    \begin{align}
        P_{\mathrm{post}} = R^2\eta_m^2+(1-\eta_m)^2-2R\eta_m(1-\eta_m)\cos\theta_n\cos\theta.
    \end{align}
    The filter-induced loss is 
    \begin{align}
        \Delta P &= P_{\mathrm{pre}}-P_{\mathrm{post}}\nonumber\\
        &=\eta_m(1- \eta_m)\left(R^2+1+2R\cos\theta_n\cos\theta\right)
    \end{align}
    with $r_n = 1$. This loss is not only set by the overlap $\eta_m$ but also by the interference controlled by $\theta_n$ through $\tilde r_s$. It is minimized with $\partial(\Delta P)/\partial \theta_n=0$ at $\theta_n=\pi$, with 
    \begin{align}
        \Delta P^{\mathrm{min}} =\eta_m(1- \eta_m)(R^2+1-2R\cos\theta).
    \end{align}
    In the large detuning limit, the $R\approx 1$ and $\theta\rightarrow0$, making $\Delta P^{\mathrm{min}}\rightarrow 0$, indicating a full suppression of the photon loss induced by mode mismatch. In particular, the conjugate symmetry $\tilde r_1=\tilde r_0^*$ is still preserved, and the protocol conditions for multi-round phase encoding can be imposed directly on $\tilde r_s$. 
        
    In Fig.~\ref{fig_SI_mismatch_I}(a,b), we show all three reflection amplitudes $|\tilde{r}_0|$, with different $\eta_m$. And the corrected amplitude $|\tilde{r}_0^{\mathrm{sel}}|$ is very close to the ideal reflection $|r_0|$ with $\eta_m=1$, demonstrating the effectiveness and releasing the requirement of the high mode overlap $\eta_m$.
    And Fig.~\ref{fig_SI_mismatch_I}(c,d) are the rest results for the repetitions $N$ and infidelity $1-F_A$ for the supplement to Fig.~4(c) in the main text. 
    
\begin{figure}[t]
        \centering
        \includegraphics[width=0.48\textwidth]{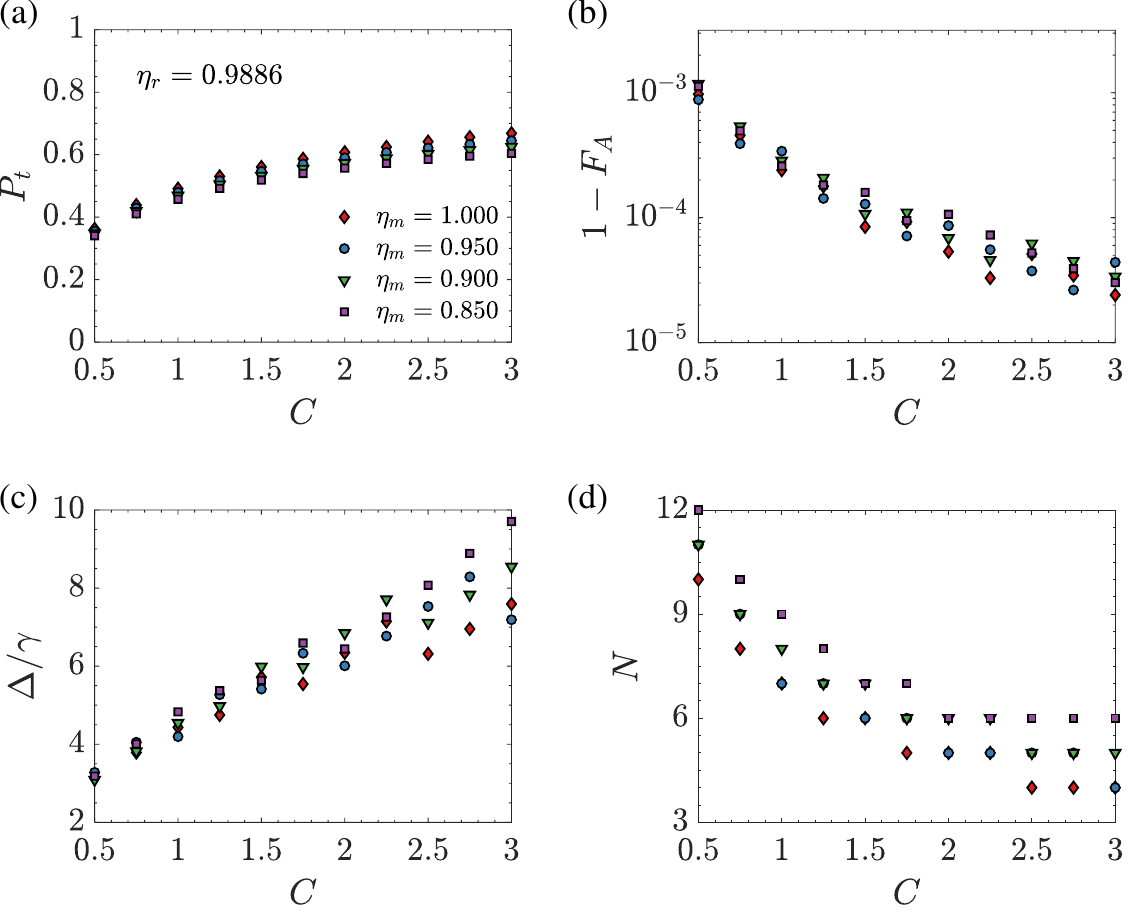}
        \caption{(a, b) The simulated maximal probability $P_t$, fidelity $F_A$, optimal values $\Delta/\gamma$ and repetition $N$ for different cooperativity $C\in[0.5,~3]$ with $\eta_r=0.9886$ and $\eta_i=0.99$. The results for different overlap $\eta_m=0.85,0.90,0.95,1$ are represented by red diamonds, blue circles, green triangles and purple squares, respectively.}
        \label{fig_SI_mismatch_II}
    \end{figure}

    Furthermore, considering the two loss channels we discussed in the last section, the overall effective reflection~\eqref{eq_SI_r_im} in large detuning  reads
    \begin{align}
        \tilde{r}_s \approx \sqrt{\eta_r}\left[ \eta_m\eta_i^2 r_s + (1-\eta_m) r_n e^{i\theta_n} \right].
    \end{align}
    Fig.~\ref{fig_SI_mismatch_II}(a,b) show the simulated maximal probability $P_t$ and the entangling fidelity with the optimal repetition $N$ as a function of cooperativity, with selective $\pi$-phase compensation applied. In this simulation, we fix two parameters $\eta_r=0.9886$ and $\eta_i=0.99$. The results for different overlap $\eta_m=0.85$, $0.9$, $0.95$, and $1$ are represented by the red diamonds, blue circles, green triangles, and purple squares, respectively. Here, both the success probability and the fidelity remain close to the ideal case ($\eta_m=1$), demonstrating that the compensation effectively mitigates mode mismatch. The corresponding optimal detuning $\Delta$ and repetition number $N$ are shown in Fig.~\ref{fig_SI_mismatch_II}(c,d) using the same markers. 
    
    Experimentally, in the first point, improving the cavity-fiber coupling overlap $\eta_m$ is one of the most important and straightforward working paths to increase the entangling probability. Recent approaches focus on engineering the optical mode delivered to the cavity so as to maximize the spatial overlap with the cavity TM$_{00}$ mode. One route employs thermal expanded-core fibers, where the controlled heating induces dopant diffusion to reduce the index difference between the fiber core and cladding. This enlarges the guided mode-field diameter, thereby enhancing the fundamental-mode overlap estimated of $\eta_m= 0.95$~\cite{Gao2019}. 
    Another strategy integrates an assembly of a graded-index and large core multimode fiber directly spliced to the single mode fiber. This all-fiber assembly transforms the propagating mode of the input single-mode fiber to match the mode of the cavity, with a mode matching of 0.9 measured~\cite{Gulati2017}. Moreover, metalenses—ultrathin metasurfaces that shape optical phase like refractive optics—have been proposed for beam shaping, e.g., by placing a metalens near a cavity mirror to mode-match a cavity to a single-mode fiber with a $\eta_m=0.94$ is expected~\cite{Wang2025}. Overall, the key improvements come from deterministically reshaping the fundamental Gaussian mode (waist size/position and wavefront curvature) to better overlap with the cavity mode. 
    
    In the second point, filtering the unmatched mode requires the ability to separate the two spatial modes. Experimentally, this can be achieved with a variety of mode-selective components, such as mode routers, optical lanterns, spatial-mode multiplexers, or single-mode filtering~\cite{Fontaine2022, Mojaver2024}; in practice, these elements must be engineered for low insertion loss to avoid excessive photon loss. Notably, in transmission-based protocols, the cavity itself acts as a mode router, transmitting the matched mode while reflecting the unmatched mode, thereby directing the two components into different paths naturally (see Sec.~\ref{Sec_SI_transmission}). 
    Finally, after separation, a relative $\pi$ phase can be implemented by adding a $\pi$ phase shift to either matched $a^\dagger_m$ and unmatched $a^\dagger_n$ modes before recombination. Alternatively, one may exploit mode-dependent phase accumulation in suitable optical components or propagation media, e.g., the differential propagation phase in a specially designed few-mode fiber. Moreover, techniques that apply a relative $\pi$ phase directly in the mode space without explicit mode separation could further facilitate an experimental implementation and have broad applications in quantum information processing~\cite{Goel2025}. This work focuses on the remote entanglement-generation protocol. A detailed optimization and implementing of such selective mode-engineering elements lies beyond the scope of this work and is better addressed by ongoing advances in photonic engineering and optical component development.
    
    \section{Three-level register}
    \label{Sec_SI_three}
\begin{figure}[t]
        \centering
        \includegraphics[width=0.485\textwidth]{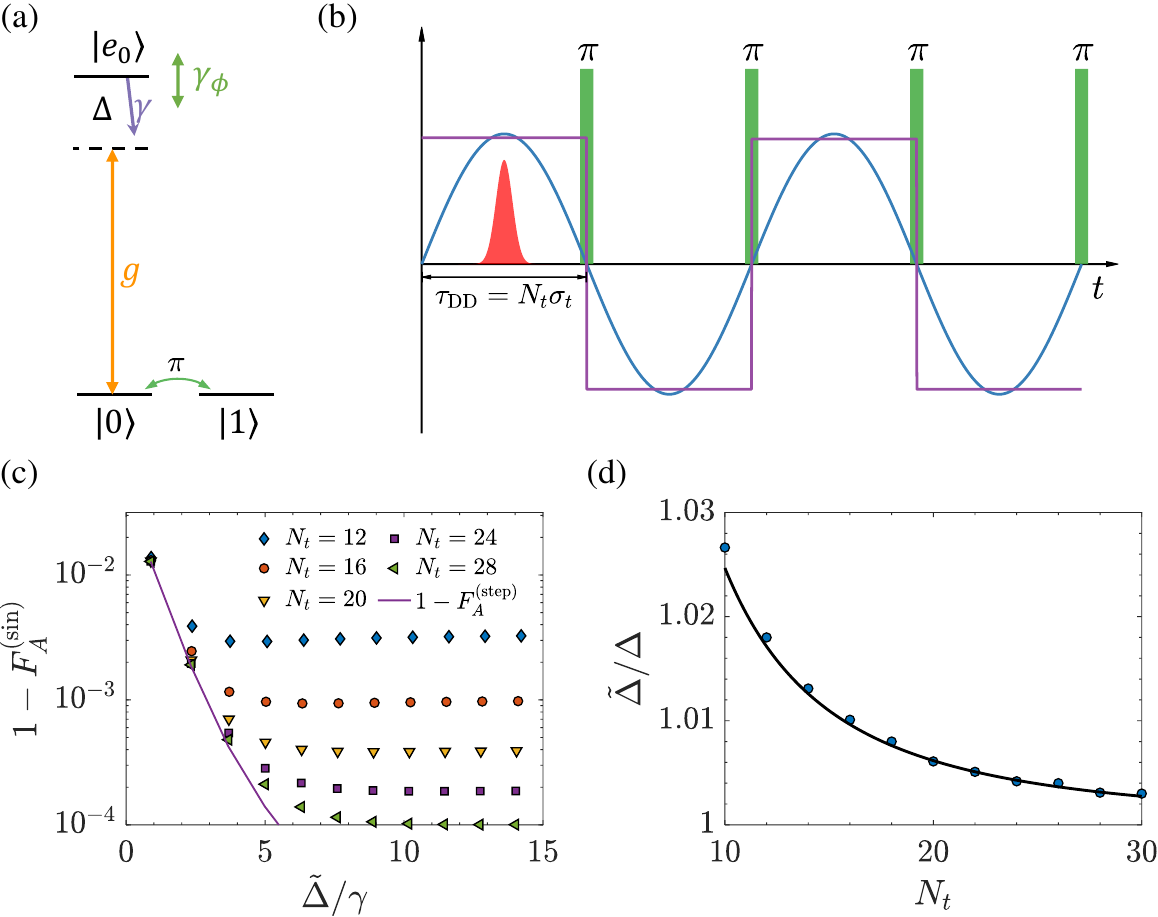}
        \caption{(a) A three-level $L$-type register: only $\mathcal{T}_0$ is coupled to the cavity.
        (b) A dynamical-decoupling-based phase-encoding scheme. The $\pi$ pulses (orange) act on the register transition $|0\rangle \leftrightarrow |1\rangle$ and flip the populations of $|0\rangle$ and $|1\rangle$, thereby reversing the sign of the accumulated phase associated with the two logical states. The detuning $\Delta(t)$ is externally controlled and alternates between $\pm\tilde{\Delta}$ in a stepwise manner (purple line), producing a piecewise-constant frequency shift that is synchronized with the pulse pattern.
        As an alternative, $\Delta(t)$ is smoothly modulated according to $\Delta(t)=\tilde{\Delta}\sin(\nu t)$ (blue), where $\nu$ sets the modulation frequency and $\tilde{\Delta}$ is the modulation amplitude.
        The incident single-photon envelope has a Gaussian temporal envelope (red) and is aligned to the sequence such that its center lies within the first half-period of the control cycle, with $\tau_{\mathrm{DD}} = N_t\sigma_t$. (c) The simulated infidelities $1-F_A^{(\mathrm{\sin})}$ versus the shifted optimal detuning amplitudes $\tilde{\Delta}/\gamma$ for sinusoidal modulation with $N_t=12,16,20,24,28$.
        The green left-pointing triangles indicate the result with a step-function reference at $N_t=10$, which approaches the results presented by the green triangles in ~Fig.~\ref{fig_simulation}(d) in the main text. (d) The corresponding modified $\tilde{\Delta}$. The black line indicates the estimated value given by~\eqref{eq_SI_tilde_Delta}.}
        \label{fig_SI_three_level}
    \end{figure}

    Our protocol can be extended to a three-level $L$-type involving register, as shown in Fig.~\ref{fig_SI_three_level}(b)~\cite{Omlor2025, Nemoto2014}. Only the transition $\mathcal{T}_0$ is coupled to the cavity, while $\mathcal{T}_1$ is far detuned and behaves as an empty cavity with reflection $r_{\mathrm{off}}(\omega)=1-2\kappa/(\kappa-2i\omega)$, which is achieved by setting $g=0$ in Eq.~(\ref{eq_r_i}). 
    A dynamical decoupling sequence of $2N$ $\pi$ pulses, depicted as green blocks in Fig.~\ref{fig_SI_three_level}(a), is applied to periodically flip $|0\rangle$ and $|1\rangle$.
    At the same instants when the $\pi$ pulses are applied, the detuning $\Delta(t)$ between $\mathcal{T}_0$ and the cavity is switched in a step-like function between $+\Delta$ and $-\Delta$, as shown by the purple line, thereby realizing the associated reflections identical to $r_0(\omega)$ and $r_1(\omega)$ in Eq.~(\ref{eq_r_i}), respectively. Consequently, for a Gaussian input pulse $\tilde{u}(\omega)$, the effective conditional modulation imparted after $2N$ rounds of interaction on a single register is given by $\left[|0\rangle\langle0|\otimes(\tilde{r}_0(\omega)^N + |1\rangle\langle1|\otimes \tilde{r}_1(\omega)^N\right] \tilde{u}(\omega)$ with $\tilde{r}_0(\omega)=r_0(\omega)r_{\mathrm{off}}(\omega)$ and $\tilde{r}_1(\omega)=r_{\mathrm{off}}(\omega)r_1(\omega)$. 
    For a monochromatic pulse, $\tilde{r}_s(0)$ reducing back to $r_s(0)$ with global phase $(-1)^N$ induced by $r_{\mathrm{off}}(0)$ are removed, rendering this equivalent to the four-level case, with only repetitions is doubled.
    
    For a Gaussian input, similarly, the entangling window appears around $\omega=0$ in which the reflections satisfy $|\tilde{r}_0(\omega)|^2\approx|\tilde{r}_1(\omega)|^2$ by construction, while tuning $\Delta$ to enforce $\tilde{r}_+(\omega)\approx 0$ in this region sets their relative phase to $\pi$.
    For improved experimental compatibility, the step-like $\Delta(t)$ can be replaced by a continuous sinusoidal $\Delta(t) = \tilde{\Delta}\sin(\nu t)$ (denoted by blue line), where $\nu=\pi/\tau_{\mathrm{DD}}$ is chosen such that $\tau_{\mathrm{DD}}=N_t\sigma_t$, and $\sigma_t = 1/\sigma_\omega$ is the temporal width of $u(t) = (\pi\sigma_t^2)^{-1/4}\exp\left(-t^2/2\sigma_t^2\right)$.
    As illustrated in Fig.~\ref{fig_SI_three_level}(a). The red curve indicates the input Gaussian pulse $u(t)$, which is precisely centered at the first segment. Regarding $u(t)$ as a weak drive~\cite{Borges2016}, the evolution of the single-excitation state $|\Psi_s(t)\rangle = c_{s,1}(t) |s,1_c\rangle + c_{e_s,0}(t) |e_s,0_c\rangle$
    is 
    \begin{align}
        \dot{c}_{s,1} &= -({\kappa}/{2}) c_{s,1} -i g c_{e_s,0} - \sqrt{\kappa_l} u(t) \nonumber \\    
        \dot{c}_{e_s,0} &= -i\left[\Delta(t) - i{\gamma}/{2}\right] c_{e_s,0} -i g c_{s,1},
    \end{align}
    with the initial conditions are $c_{s,1}(0) = c_{e_s,0}(0) =0$ and $|k_c\rangle$ is the Fock state of cavity mode. According to input-output relations~\cite{Agarwal2013, Collett1984, Gardiner1985, Reiserer2015}, the temporal shape of output photon after the first round is given as $u_{\mathrm{l},s}(t) = u(t) + \sqrt{\kappa_l} c_{s,1}(t)$, and then $u_{\mathrm{l,s}}(t)$ will be regarded as a new input state and sent back to interaction again and repeated until the final round. Then the final state numerically in single register as $\int dt \left[|0\rangle u_{0}(t) +|1\rangle u_{1}(t) \right]\hat{a}^\dagger(t)|\mathrm{vac}\rangle$, resulting in probability 
    \begin{align}
        P_A &= \int dt\left[2|u_+(t)|^2+|u_-(t)|^2\right]/2,\nonumber\\
        P_B &=\int dt|u_-(t)|^2/2
    \end{align}
    with the corresponding fidelities given $F_A^{(\sin)}=P_B/P_A$, $F_B^{(\sin)}=1$. Here $u_s(t)$ is the final temporal wave packet of the reflected photon associated with spin state $|s\rangle$, and $2u_{\pm}(t) = u_0(t)\pm u_{1}(t)$.
    In Fig.~\ref{fig_SI_three_level}(c), we show the infidelities $1-F_A^{(\sin)}$ as functions of $N_t$ for $C=2$ with the optimally chosen $\tilde{\Delta}$. The green diamonds indicate $1-F_A^{(\mathrm{step})}$ obtained with a stepwise $\Delta(t)$ at $N_t=10$. For larger $N_t$ with the sinusoidal $\Delta(t)$, the fidelity $F_A$ exceeds $0.99$ and continues to improve while still being lower bound with the $1-F_A^{(\mathrm{step})}$. We note that the corresponding success probabilities $P_t$ remain essentially unchanged.
    
    When the Gaussian pulse is centered near a detuning maximum $t_c$ with $\nu t_c=\pi/2$, the detuning sampled by the pulse is slightly reduced because $\sin(\nu t)$ is curved around its maximum. Expanding around $t_c$ gives
    \begin{align}
        \Delta(t)&=\tilde{\Delta}\sin(\nu t)=\tilde{\Delta}\cos(\nu(t-t_c))\nonumber\\
        &\approx \tilde{\Delta}\left(1-\frac{\nu^2(t-t_c)^2}{2}\right).
    \end{align}
    Here, we perform an intensity-weighted average, since the relevant figures of merit (mode overlap and success probabilities entering the fidelity) are quadratic functionals of the field. For a Gaussian envelope centered at $t_c$ we use $\langle (t-t_c)^2\rangle\sim \sigma_t^2/2$ which yields the effective detuning amplitude sampled by the pulse
    \begin{align}
        \Delta_{\mathrm{eff}}\approx \tilde{\Delta}\left(1-\frac{\nu^2\sigma_t^2}{4}\right).
    \end{align}
    We obtain the simple scaling
    \begin{align}
        \frac{\tilde{\Delta}}{\Delta}\approx 1+\frac{\pi^2}{4N_t^2}.
        \label{eq_SI_tilde_Delta}
    \end{align}
    It is indicated by the black line in Fig.~\ref{fig_SI_three_level}(d), which agrees well with the values numerically obtained.
    
    \section{The transmission-based protocol}
    \label{Sec_SI_transmission}
    In this section, we will briefly introduce the entangling protocol using the transmitted photon as the resource. Here, unlike the reflection-based one, a symmetric cavity with $\kappa_l=\kappa_r$ is employed to ensure a transmission~\eqref{Eq_SI_rt} approaches $-1$ in the large detuning. Considering a monochromatic pulse, the transmission is
    \begin{align}
        t_0=-1-i\frac{C}{O} + \frac{C(1+C)}{O^2}+\mathcal{O}\left(\frac{1}{O^3}\right)\approx -Te^{i\theta_t}
    \end{align}
    and $t_0=t_1^*$.
    Compared to the results of reflection, we have 
    \begin{align}
        R\approx2T,\quad \theta\approx2\theta_t,
    \end{align}
    thereby resulting in a double repetition required. This longer duration is the main disadvantage of the transmission-based protocol.
    Similarly, for the Gaussian-shaped input, the entangling condition is $t_{+}(\omega)=0$, with $2t_{\pm}(\omega)=t_0(\omega)\pm t_1(\omega)$, and the total success probability is $\int d\omega |t_{-}(\omega)|^2|\tilde{u}(\omega)|^2$.  
    In the following, we will analyze its performance as we have done for the reflection one, only under the monochromatic limit.
    
    As for the imperfections lowering the probability, e.g., cavity loss, optical loss, mode mismatching between the fiber and cavity, we could write the modified transmission as 
    \begin{align}
        \sqrt{\eta_r}\eta_i \eta_m t_s(\omega).
    \end{align}
    Here $\eta_r$ is the optical loss of the circuit. We note that only one $2\times2$ optical switch is needed, thus half of the losses. Considering the intrinsic loss $\kappa_i$, the transmission is 
    \begin{align}
        t_s(\eta_i) = \frac{-2\kappa\eta_i(\gamma+2i\Delta_s)}{\kappa(\gamma+2i\Delta_s)+4g^2},
    \end{align}
    giving that 
    \begin{align}
        t_s(\eta_i)/t_s(\eta_i=1) = \eta_i.
    \end{align}
    Moreover, as for the mode mismatching, the matched mode $a^\dagger_m$ enters the cavity and is transmitted to the right side of the cavity. However, the unmatched mode $a^\dagger_n$ will be directly reflected. Thus, the matched and unmatched modes will naturally be routed to two different paths, be equal to filtering the unmatched mode. This is similar to the case with respect to the mode filtering correction of reflection in Sec.~\ref{Sec_mode_mismatch}.

    With respect to the differences may appear inside the register $R_M$, similarly, we could solve the equation $|t_{0,M}|^2=|t_{1,M}|^2$ to obtain the solution of the detuning 
    \begin{align}
        \Delta_{1,M}= - \frac{\gamma_{1,M}}{2}\sqrt{\frac{|t_{0,M}|^2(1+C_{1,M})^2-1}{1-|t_{0,M}|^2}}.
    \end{align}
    We could increase the detuning $\Delta_{0,M}$ to make $|t_{0,M}|^2(1+C_{1,M})^2-1>0$ thereby solving the value of $\Delta_{1,M}$.
\end{document}